\documentclass{sig-alternate}
\usepackage[hidelinks, pdfpagelabels=false]{hyperref}
\usepackage{graphicx}
\usepackage{listings}
\usepackage{color}
\usepackage{float}
\usepackage{array}
\newcolumntype{L}[1]{>{\raggedright\let\newline\\\arraybackslash\hspace{0pt}}m{#1}}
\newcolumntype{C}[1]{>{\centering\let\newline\\\arraybackslash\hspace{0pt}}m{#1}}
\newcolumntype{R}[1]{>{\raggedleft\let\newline\\\arraybackslash\hspace{0pt}}m{#1}}
\usepackage{threeparttable}
\usepackage{etoolbox}
\appto\TPTnoteSettings{\footnotesize}

\floatstyle{plaintop}
\newfloat{program}{thp}{lop}
\floatname{program}{Program}

\begin{document} \sloppy

\setlength{\footnotesep}{0.4cm}

\definecolor{mygreen}{rgb}{0,0.6,0}
\definecolor{mygray}{rgb}{0.5,0.5,0.5}
\definecolor{mymauve}{rgb}{0.58,0,0.82}

\lstset{ %
  backgroundcolor=\color{white},   % choose the background color
  basicstyle=\scriptsize,          % size of fonts used for the code
  breaklines=true,                 % automatic line breaking only at whitespace
  captionpos=b,                    % sets the caption-position to bottom
  commentstyle=\color{mygreen},    % comment style
  escapeinside={\%*}{*)},          % if you want to add LaTeX within your code
  keywordstyle=\color{blue},       % keyword style
  stringstyle=\color{mymauve},     % string literal style
}

\conferenceinfo{Proceedings of the C++ Now 2014 Conference}{Aspen, Colorado, USA.}
\CopyrightYear{2014}
\crdata{Niall Douglas}

\title{Large Code Base Change Ripple Management in C++}
\subtitle{My thoughts on how a new Boost C++ Library could help
  \numberofauthors{1}
  \author{
  \alignauthor
  Niall Douglas\\
       \affaddr{Waterloo Institute for Complexity and Innovation (WICI)}\\
       \affaddr{University of Waterloo}\\
       \affaddr{Ontario, Canada}\\
       \email{\url{http://www.nedprod.com/}}
  }
}
\maketitle
\pagenumbering{arabic}
\begin{abstract}
C++ 98/03 already has a reputation for overwhelming complexity compared to other programming languages. The raft of new features in C++ 11/14 suggests that the complexity in the next generation of C++ code bases will overwhelm still further. The planned C++ 17 will probably worsen matters in ways difficult to presently imagine.

Countervailing against this rise in software complexity is the hard de-exponentialisation of computer hardware capacity growth expected no later than 2020, and which will have even harder to imagine consequences on all computer software. WG21 C++ 17 study groups SG2 (Modules), SG7 (Reflection), SG8 (Concepts), and to a lesser extent SG10 (Feature Test) and SG12 (Undefined Behaviour), are all fundamentally about significantly improving complexity management in C++ 17, yet WG21's significant work on improving C++ complexity management is rarely mentioned explicitly.

This presentation pitches a novel implementation solution for some of these complexity scaling problems, tying together SG2 and SG7 with parts of SG3 (Filesystem): a standardised but very lightweight transactional graph database based on Boost.ASIO, Boost.AFIO and Boost.Graph at the very core of the C++ runtime, making future C++ codebases considerably more tractable and affordable to all users of C++. 
\end{abstract}



\section{Introduction}
This paper ended up growing to become thrice the size originally planned, and it has become rather unfortunately information dense, as each of the paper's reviewers asked for additional supporting evidence in the wide field of topics touched upon -- extraordinary claims require extraordinary evidence after all. It does not do justice to each of those topics, and I may split it next year into two or three separate papers depending on reception at C++ Now.

For now, be aware that I am proposing a new Boost library which provides a generic, freeform set of low level content-addressable database-ish layers which use a similar storage algorithm to git. These can be combined in various ways to create any combination of as much or as little `database' as needed for an optimal solution, including making the database active and self-bootstrapping which I then propose as a good way of extending SG2 Modules and SG7 Reflection in C++ 17 to their maximal potentials, specifically to make possible `as if all code is compiled header only' including non-C++ code in a post-C++ 17 language implementation. This paper will -- fairly laboriously -- go through, with rationales, all the parts of large scale C++ usage both now and my estimations of during the next decade which would be, in my opinion, transformed greatly for the better, in an attempt to persuade you that supporting the development of such a new Boost library would be valuable to all users of C and C++, especially those programming languages which still refuse to move from C to C++ due to C++'s continuing scalability problems. Donations of time, money, equipment or design reviewing eyeballs are all welcomed, and you will find more details on exactly what is needed in the Conclusion.

\tableofcontents

\section{What makes code changes ripple differently in C++ to other languages?}
How difficult is writing code in C++ compared to the other major programming languages? A quick search of Google shows that this is a surprisingly often asked question: most answers say it is about as hard as Java, to which probably most at this conference will chortle loudly. However, probably for the vast majority of C++ programmers who never stray much beyond Qt-level mid-1990s C++ (let us call this `traditional C++' from now on), that complexity of C++ is about par with a similar depth into Java, or for that matter Smalltalk, C\# or any object-orientated programming language. As a human language analogy, from a skin deep level object-orientated languages look just as similar as Spanish, French, Italian and Latin do on first glance.

Yet with experience traditional C++ programmers start to notice some odd things about C++ once you start writing and maintaining some moderately large C++ code bases as compared to writing and maintaining moderately large code bases written in Java -- and I'll rank these in an order of increasing severity according to my personal opinion:

\begin{enumerate}
\item \textbf{Undefined behaviours}: unlike Java which has a canonical implementation, the majority, but not all, of C++ is specified by an ISO standard which is implemented, with varying quirks and bugs (`dialects'), by many vendors. Which vendor's implementation is superior isn't as relevant in practice as which C++ dialect is compatible with the libraries you're going to be using, and it's not unheard of for some commercial libraries to quite literally only compile with one very specific compiler and version of that compiler. This is fine, of course, until you try to mash up one set of libraries which require specific compiler and runtime A with another set of libraries which require specific compiler and runtime B, which can result in having to write wrapper thunk functions in C for every API mashed up (fun!).

\item \textbf{The fragile binary interface problem} i.e. mixing up into the same process space binaries of different versions of libraries, or binaries compiled with non-identical STLs is dangerous: while compiler vendors go to great lengths to ensure version Y of their compiler will understand binaries compiled by version X of their compiler (where X\textless Y), the same is not so true for the standard library runtimes supplied across compiler versions e.g. woe betide you should you try linking against a library compiled against \texttt{libstdc++.so.5} into a runtime using \texttt{libstdc++.so.6} -- similarly, bringing a DLL linked against \texttt{MSVCRT80.DLL} into a process full of \texttt{MSVCRT100.DLL} linked DLLs is likely to not be entirely reliable\footnote{Though plenty of people do it anyway of course, and then \hyperlink{http://git.postgresql.org/gitweb/?p=postgresql.git;a=blob;f=src/port/win32env.c;hb=HEAD}{code such as this evil in PostgreSQL becomes necessary} (\url{http://git.postgresql.org/gitweb/?p=postgresql.git;a=blob;f=src/port/win32env.c;hb=HEAD}).}. The unavoidability of mixing library versions in a single process is at its worst with the STL as it's the most commonly used C++ library of all, but it is a problem with any commonly used C++ library due to the inability to control what your dependant libraries do, which affects everything from Qt to Boost.

\item \textbf{The fragile base class problem}: Java's ABI is quite brittle compared to other languages such that it is too easy to accidentally break binary compatibility when you change a library API -- well, the C++ ABI is far more brittle again due to (i) using caller rather than callee class instantiation, thus making all class-\emph{using} code hardwired to the definition of that class and (ii) using offset based addressing rather than signatures for both virtual member functions and object instance data layouts, which let you too easily break binary compatibility without having any idea you have done so until your application suddenly starts to randomly segfault. The PIMPL idiom of hiding data layouts in a private class and defining all public classes to have an identical data layout (a single pointer to the opaque implementation class) is usually recommended at this stage (and is heavily used by Qt and Qt-like codebases), but PIMPL has significant hard performance overheads: it brings in lots of totally unnecessary calls to the memory allocator for stack allocated objects, and it actively gets in the way of code optimisation and reflection. More importantly for this paper's purposes, \emph{PIMPL does not work well with the idioms and constructs made possible by C++11/14 and later} -- think to yourself how you would go about assembling a PIMPL idiom with template metaprogramming for example\footnote{To a limited extent it can be done with lots of hoop jumping -- for example proposed Boost.AFIO does so.}?

\item \textbf{Memory corruption}: a single piece of bad code buried in some minor library hidden away in some dependency representing a fraction of a percentage of total line count can utterly ruin the quality and reliability of everything else in that process address space. I think this `feature' of C++ (and to a lesser extent, C) is probably unique to C++ out of all modern programming languages actually -- it is certainly unknown to Java programmers, most of whom become very unstuck if transferred to work in large scale C++ development with insufficient training. Some may not think this too much of a problem: however, it is a \emph{huge} problem if you are forced to use third party supplied binary blobs which you \emph{know} are corrupting memory, and the vendor in question refuses to do anything about it even after you have incontrovertibly proved the problem. I personally have not yet worked in a corporate position where this has not happened with third party supplied binary blobs, sadly it is amazingly common.

\item \textbf{State corruption}: worse than memory corruption is \emph{state corruption} which is generally, but not always, caused uniquely in C++ rather than in other languages by two main causes: (i) race conditions generated by more than one thread of code modifying memory without being written to handle it correctly (this includes the direct use of condition variables without wait predicates, which is scarily common despite that very little such code is race free\footnote{See the new Boost permit object for a safe drop-in replacement for unsafe condition variable usage.}) and (ii) exceptions being thrown up the call stack through C++ which is not exception safe (even if written to be exception safe). The reason that state corruption is worse than memory corruption is because memory corruption detection tools such as \texttt{valgrind} will not detect state corruption for you. State corruption due to thread race conditions can be detected using tools such as the clang/GCC \texttt{ThreadSanitizer} or helgrind/DRD tools in \texttt{valgrind}, however you are quite on your own when trying to debug exception safety breakages, and there I sadly know of no better tool than the venerable \texttt{printf()}. As much as state corruption is a serious issue in \emph{any} of the modern non-functional programming languages, most have not retrofitted threading and structured exception handling to the language long after there were already many tens of million line C++ projects in existence -- and then gone on to witness widely used C++ codebases and libraries to intentionally and deliberately not retrofit their code to support either exceptions (still surprisingly common) nor threading (less common), which has led to \emph{incommensurability problems} in C++ codebases.

Incommensurability is a term borrowed from post-structuralism, but I mean it in the same sense as Kuhn's seminal 1962 book \cite{kuhn1962structure}. An illustrative example might be when people use the STL in Qt code: what happens if STL code throws an exception in a Qt slot implementation? Answer: Qt fatal exits the process because state has become undefinable. Given that \texttt{operator new} is technically part of the STL, does this not mean that all Qt-based modern C++ cannot be considered to have well defined state if it is \emph{ever} possible for the system to run out of memory? Answer: correct, because you cannot allow any situation where an exception might enter Qt code, otherwise known state is permanently lost. What therefore happens when you mix exception safe code and exception unsafe code? Answer: all code must now be considered exception unsafe -- \emph{just one line} of exception unsafe code is enough to theoretically taint the exception safety of every other line of code in the process, which is a pretty extreme outcome.
\end{enumerate}

All these C++-peculiar issues have turned up throughout my career, and I would assume that the same issues are very familiar to those at this conference as well. As much as they are irritating in a moderately large codebase, they are usually not too fatal because there is one almighty hack to work around most of their manifestations: \textbf{always recompile everything potentially affected by a source code change}, with the further reification of \textbf{make all implementation possible header-only} i.e. always recompile everything \emph{per-compiland} which is of course a favourite of Boost where ten minute per compiland compile times are entirely possible. That eliminates versioning and ABI brittleness problems by definition, and therefore lets you use all the modern C++ idioms and techniques you like without having to worry about the terrible consequences on binary stability.

I think it self-evident to all at this conference that the use of those modern techniques lets you substantially drive down the amount of memory corruption and the cost of fixing memory corruption, and I suspect it has highly positive effects on state corruption as well, not least that newer code is invariably better tested and better designed to handle threads and exceptions than older code was at the same stage of maturity. The costs are mainly on the skill level required by the engineers and the very ample CPU resources required to repeatedly compile so much code.

\section{Why hasn't C++ replaced C in most newly written code?}

\begin{figure}[t]
\centering
\includegraphics[width=85mm]{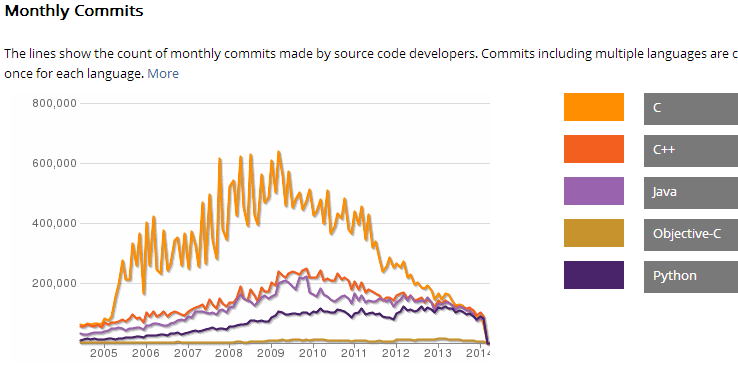}
\caption{Monthly commits to open source projects tracked by Ohloh.net by programming language 2004-2014.}
\label{CvsCPP1}
\end{figure}

\begin{figure}[t]
\centering
\includegraphics[width=85mm]{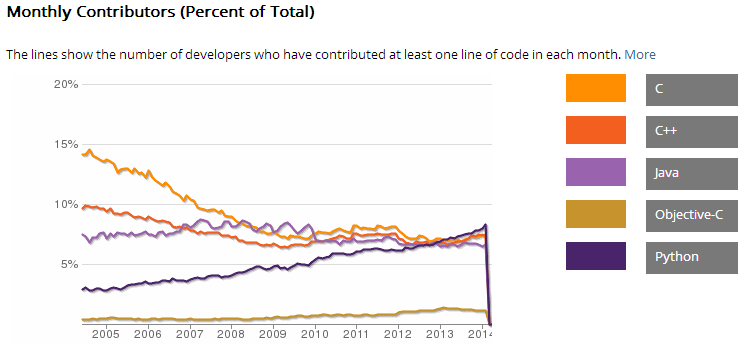}
\caption{Monthly contributors to open source projects tracked by Ohloh.net by programming language as percentage of total 2004-2014.}
\label{CvsCPP2}
\end{figure}

Given that Objective C and C++ are the two most successful attempts at creating a better C whilst remaining compatible with C, I am going to ask a question rarely asked nowadays: why is it that in 2014 there is still more new code written in C than C++\footnote{Chandler Carruth last night raised the question with me: what do I mean here by `new code'? By new code I mean code newly altered, so that includes bug fixes in old code bases as well as greenfield code.}?

In Figures \ref{CvsCPP1} \& \ref{CvsCPP2} you will see the number of open source commits and developers per month working in the programming languages C, C++, Objective C with Java and Python thrown in for good measure, with the effects of the recent economic downturn very plain to see. While the choice of programming language in open source code is probably not representative of the entire industry, it is probably not far off, and it illustrates some very interesting things.

Firstly, the continuing popularity of C for new code is quite remarkable (assuming that Ohloh's parsers are good at distinguishing C from C++ and Objective C). Java did, for about a year around 2009, perhaps surpass it to become the most popular language for new open source code, but the 2010 acquisition of Sun Microsystems by Oracle had a rapid chilling effect, and Java has been slowly trending downwards since.

Secondly, Python's growth appears to be unstoppable, and is now or shortly will be the most popular programming language for new open source code apart from Javascript. You'll see me mention Python a lot during the rest of this paper, mainly because I expect it will be \emph{the} mainstream applications and services programming language of the next decade, and in my opinion C++ needs to start accommodating that explicitly.

Thirdly, now is a very unusual time in that all the major programming languages have about equal shares on all metrics, where it is in number of commits, number of developers or number of active projects. I am not aware of this having ever occurred before in history, so even if just confined to new open source code, I think it safe to say a structural transition is occurring.

Which then begs the question: why hasn't C++ completely supplanted C for new code as was once expected/hoped for? Why is it that many C codebases (SQlite3, OpenSSL and GCC all pop into mind) choose to extend C with macro implementations of some C++ features instead of using C++ (note that GCC now allows C++ in its source code)? Why is it that many C codebases have virulently anti-C++ policies?

Some of it can be explained by language taste: Linus Torvalds is but one of many who dislike C++ as a language for example, but even a strong personal dislike usually doesn't forbid a whole language being used in some submodule somewhere. I think, personally, that what happens in practice is that there are several showstoppers for C++ in really large mixed language code bases:

\begin{itemize}
\item C++ extensively pollutes the C symbol namespace with lots of mangled symbols. As much as the mangling avoids any conflicts, it slows down linking considerably, looks messy, and generally provides evidence that C++ has conspicuously failed to provide formal facilities for ABI management.
\item PIMPL is in fact the state of the art in ABI management techniques, but it is an anti-optimisation, anti-transparency, anti-inspection technique which has significant hard runtime costs (especially on the memory allocator) associated with it. Looked at from outside C++, it looks like a temporary hack to workaround a failure to deal properly with ABI management.
\item I don't think it helps that all the recent action in new C++ features have been highly introverted and navel-gazing -- if you examine C++ 11/14's new features which are not also replicated in C11, they are all about dealing with C++ problems caused unto C++ by C++ itself. They do not a jot for those looking to use C++ with other languages.
\end{itemize}

Let me sum up this section as follows: if you are a Python runtime engineer and you look at what C++ 11/14 gives you which C11 does not, I don't think you will find anything at all which would make you reconsider C++ as a choice for the Python interpreter. C++ 11/14 adds a ton of great stuff, but did any of it persuade someone like Linus that C++ might be tolerable in the Linux kernel? Do the Python/Ruby/Lua/PHP interpreter guys look at C++ 11/14 and go `wow that transforms our use case for C++ over C'?

There is of course nothing wrong with looking inwards for a standards release, but I most definitely think you need to flip that focus for the next standards release, or else risk becoming a niche programming language for specialist uses instead of the end all and be all systems programming language which C++ has always aspired to be: a truly superior C, one where there can be no doubt that it is the superior choice for any new systems project under any metric.
 
\begin{figure}[t]
\centering
\includegraphics[width=85mm]{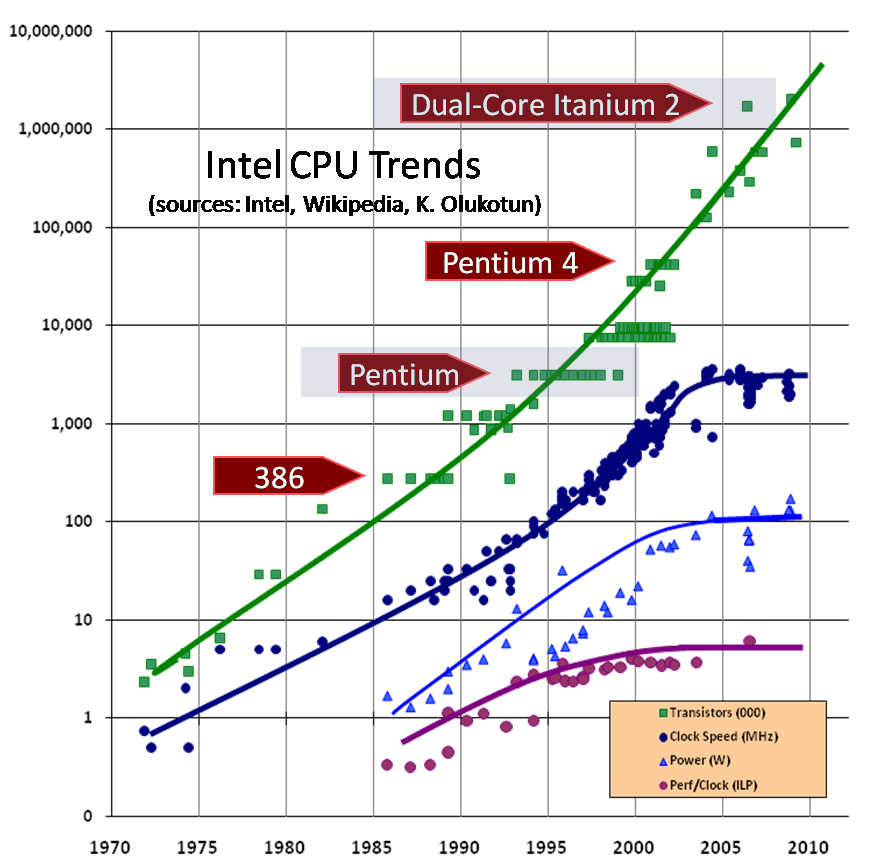}
\caption{The graph from the famous `The Free Lunch is Over' paper by Sutter (2005; this graph updated to 2009) \cite{sutter2005free}. Note how growth in clock speed went linear around 2003, but growth in transistor density remains exponential for now.}
\label{freelunchover1}
\end{figure}

\begin{figure}[t]
\centering
\includegraphics[width=85mm]{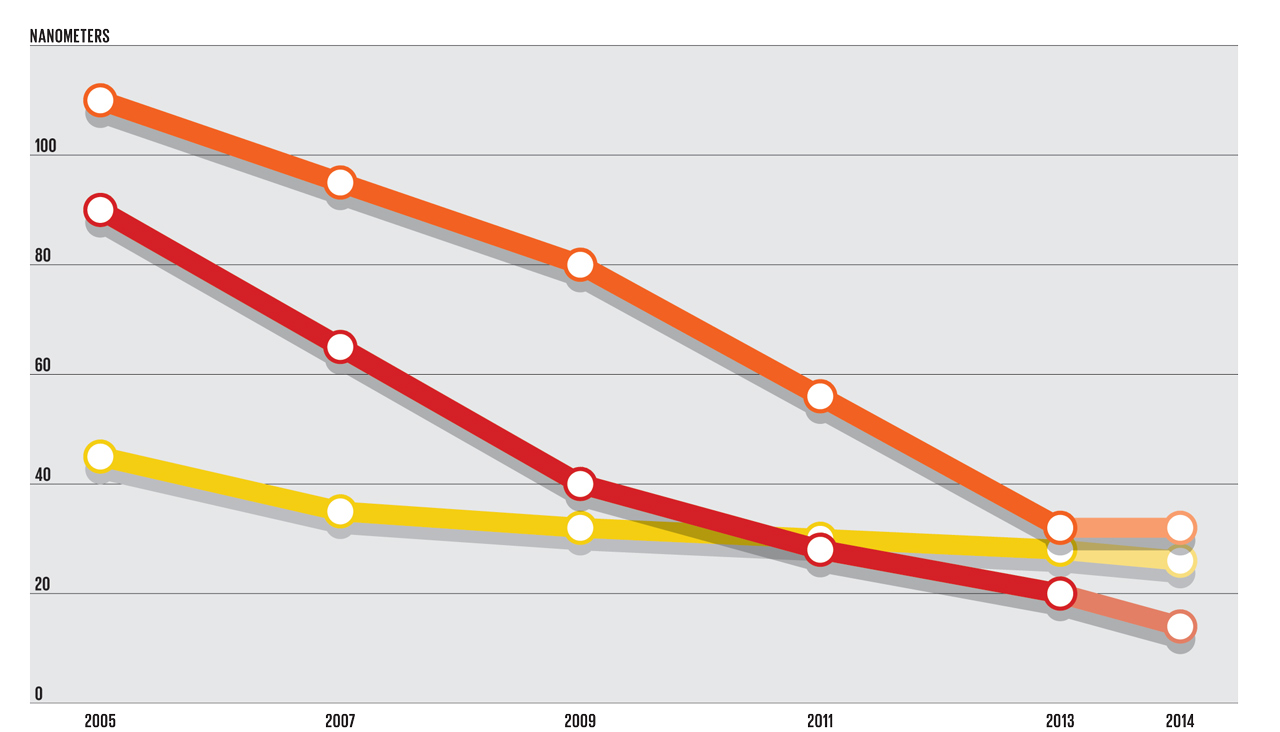}
\caption{Historical comparison of advertised process size (red line) to transistor gate length (yellow line) and the metal one half pitch (orange line) which is a measure of wiring density. Note how the latest 14 nm process actually uses the same one half pitch as the 22 nm process, the transistor density improvements came from extending transistors into three dimensions and reorganising circuitry layouts, both of which can only buy you a few more process shrinks \cite{spectrum2013}.}
\label{freelunchover2}
\end{figure}

\begin{figure}[!ht]
\centering
\includegraphics[width=85mm]{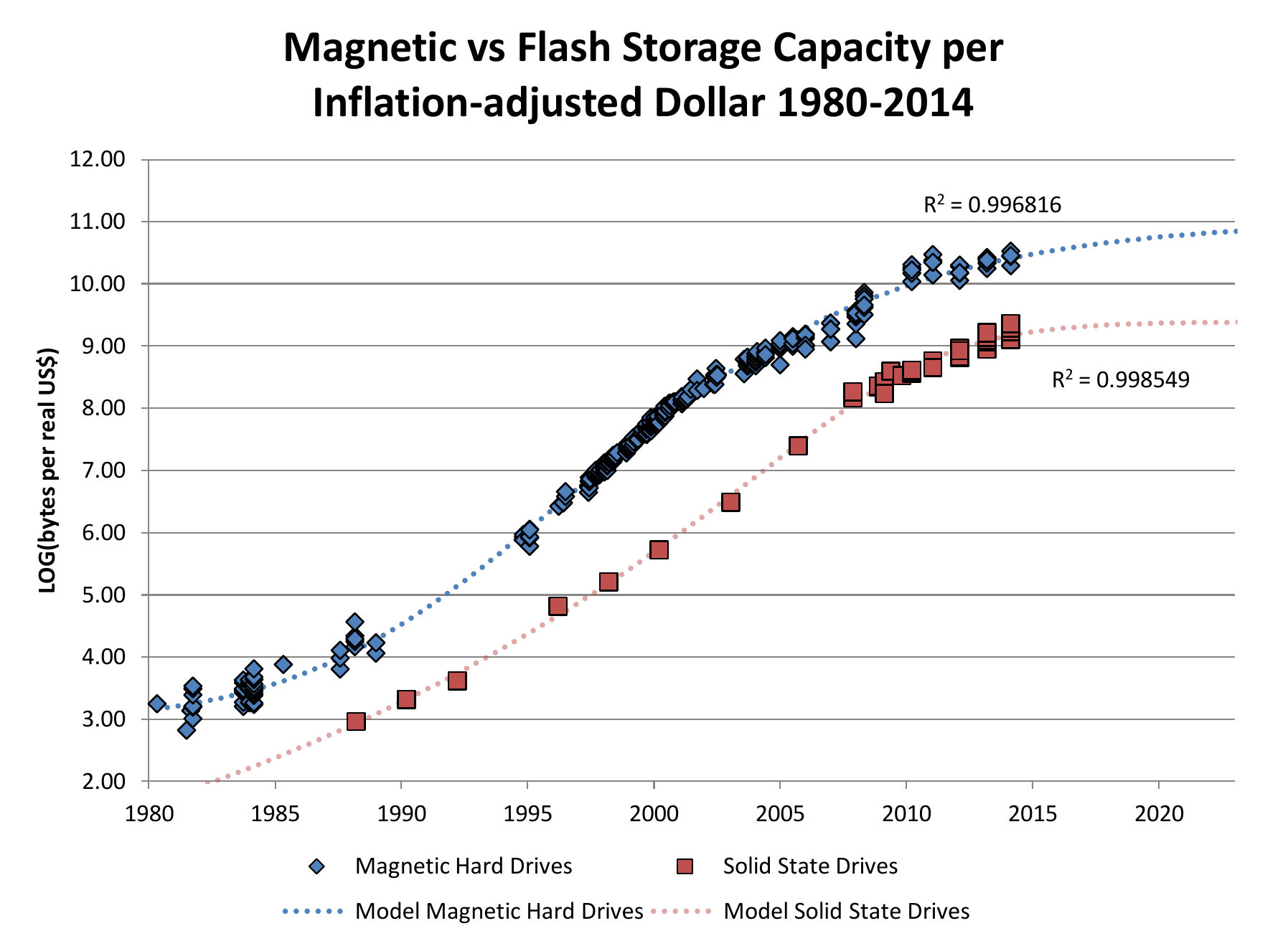}
\caption{Historical comparison of Magnetic to Flash Storage capacity per inflation-adjusted dollar. Data compiled and regressed by myself. Excel spreadsheet containing raw data can be found at \cite{SSDsVsHardDrives}.}
\label{freelunchover3}
\end{figure}

\section{What does this mean for C++ a decade from now?}

What I'm going to talk about next is necessarily somewhat speculative given that no one knows what C++ 17 will eventually define, or even how C++ 11/14 will end up becoming used by the wider practicing community. I am going to make the assumption though that we are currently living in a very unusual golden age made only possible by Moore's Law having different logistic growth curves for clock speed and transistor densities as you'll see in Figure \ref{freelunchover1}. What this means is that because build systems can parallelise compilation across available processors, compiling code is still seeing exponential year-on-year improvements per currency unit expended. Similarly, we are able to offload a lot of testing onto continuous integration servers, which similarly see exponentially falling costs per unit of testing.

I'll be frank in saying that I very much doubt that this unusual disparity will persist for many years longer with transistor density growth also going linear in the next few years (see Figure \ref{freelunchover2}), and with it will pass the unique conditions which let us lazily throw ever more C++ source code into per-compiland consideration. Combined with this de-exponentialisation in growth of transistor density will be the continuing exponential growth in data storage capacity (see Figure \ref{freelunchover3}), which means linearly improving CPUs will have to deal with exponentially rising quantities of data, something anyone who has cloned a hard drive knows all about (storage access rate growth went linear two decades ago, so copying a hard drive can take the best part of a day nowadays).

Put more simply, right now we can freely add additional C++ complexity to a codebase knowing that the costs of additions are likely to be soaked up by the exponential growth in transistor density. Come 2017-2020 however, for every one unit of additional C++ complexity added, at least one unit of additional build time and testing time will be added. This \emph{profoundly} changes the complexity economics of computer software.

I am therefore going to assert, given the trends illustrated above, that by around 2017-2020 and \emph{if we are still using present generation C++ toolsets}, we're going to see us stop including as much source code as possible via header only techniques, and not long after we're going to have to stop recompiling all potentially rather than actually affected source code\footnote{i.e. the source files that the build tool thinks could potentially be affected by a change. This is usually considerably more than is actually needed.} every time we make a single change too. On top of that, executable and dataset sizes have little impetus to stop growing as storage capacity remains growing exponentially, so we're going to keep piling on more lines of code and executables are going to keep growing BUT less will be compiled in a given compiland, and more linking of more compiled objects will be happening. In other words, I am going to predict that by around 2020, and assuming that present trends continue and that C++ 17 looks reasonably like what has been already proposed for it, C++ development in the 2020s is going to look a bit more like C++ development was in the 1990s where linking your object files into an executable was a real chore, except all those fancy new metaprogrammed features are going to stomp all over your organisation's productivity as our hack of recompiling everything every compile run becomes no longer feasible. Oh, and you'll probably be doing all this with a lot of Python code standing not too far away from your APIs and ABIs.

Of course, so far I've only been talking about moderately large code bases as these are the majority. Very large C++ code bases (i.e. the ones which take a day or more to recompile with hundreds of CPUs working on the problem) \emph{already} have these problems -- Microsoft, Google and BlackBerry all place very substantial restrictions on the extent to which their engineers can use cutting edge C++ techniques across ABI boundaries (as in, \emph{use no C++ features which weren't common before year 2000}), with extensive automated checking of each code commit for ABI breakages. In fact, all three internally recommend that you don't even throw exceptions across an ABI boundary, and exceptions entered the ISO C++ standard back in 1998! Certainly while I worked at BlackBerry including anything from Boost in a public API was highly frowned upon as Boost does not attempt to preserve a stable ABI over time, and I would doubt there is any difference at any of the other software majors. Imagine then to yourself how ABI stable capturing lambdas with their unique, and intentionally unknowable type\footnote{Unlike non-capturing lambdas which can decay to a C-style function pointer, capturing lambdas have some unique but unknown type which implements the call operator.} and therefore \emph{unknowable} ABI are, and now consider how you would write ABI stability rules for a language feature like capturing lambdas? The answer is obvious: the only option with current technologies is to ban capturing lambdas going anywhere near an ABI boundary\footnote{An obvious workaround is to type erase the capturing lambda type into a \texttt{std::function<>}, and have the ABI accept only one of those as the sole means of transporting a callable type. This, however, loses information i.e. information is erased by the ABI boundary, plus introduces overhead.}.

You can, in fact, rinse and repeat this thought exercise with many, if not most of the new language and standard library additions to the 2011 C++ standard: variadic templates are by definition anti-ABI-stability, as is anything based on them like poor friendly \texttt{std::tuple<>} which is also best banned from going near an ABI boundary.

This raises an unpleasant spectre: that an ever increasing share of C++ language features added since the mid-1990s are going to be permanently unavailable to large C++ code bases as time moves forward, with a trickle down as Moore's Law progresses into linearity into more medium sized code bases. In short, ladies and gentlemen, we are facing into what the complex systems literature would call a medium term C++ complexity scaling exigency.

\section{What C++ 17 is doing about complexity management}

None of the above problems have passed over the WG21 committee of course: WG21 C++ 17 study groups SG2 (Modules), SG7 (Reflection), SG8 (Concepts), and to a lesser extent SG10 (Feature Test) and SG12 (Undefined Behaviour), are all fundamentally about significantly improving code complexity understanding and/or management in C++ 17, albeit to my understanding from a still C++-centric perspective (rather than from say the perspective of the Python interpreter). It's worth having a quick look at some of those efforts, and what `wicked hard' problems C++ 17 has explicitly put \emph{off} the table already as issues better tackled in an even later standard.

\subsection{WG21 SG2 (Modules)}
Right now in C++ the standard doesn't say much about binary object encapsulation: basically the standard is written around the unspoken assumption that you are able (even if you don't) to compile all code entering a process' space which is probably an untenable assumption given how large the STL has become in C++ 11\footnote{By this I mean that it was doable to include a header implementation of most of a C++ 03 STL with every source file compiled. This is definitely not possible with most of a C++ 11 STL, the compiler would surely run out of memory.}. This has led to some proprietary vendor innovations, of which three have become dominant in C++: (i) the ELF shared object approach (ii) the Mach O/PE dynamically linked library (DLL) approach and (iii) the Microsoft Component Object Model (COM) approach.

The ELF shared object approach tries to be as thin an abstraction from a collection of object files produced after compiling as possible i.e. pretend you're just linking a very large single program, so keep a single global symbol table and if collisions happen, so be it; the Mach O/PE DLL approach where symbol tables are kept per-DLL, thus avoiding most but not all problems caused by collisions; while the COM approach is to eliminate as much C++ complexity as possible (only virtual member functions allowed, and no exceptions, only return codes), and provide the absolute minimum information layer to permit as loose a binding as possible, which while effective for very large scale C++ development, it also removes almost all of C++ from the ABI boundary. Note that of those three proprietary innovations, the first two are actually for C code, while the third is 100\% compatible with C code. All three are very simple in theory -- perhaps why they have been so successful -- and all three are hideously complex when you start pushing the envelope, which is why you can find books wholly dedicated to just each topic.

The SG2 Modules proposal (rev. 6) N3347 \cite{n3347}, and the only implementation of it that I am aware of which is in clang \cite{modulesinclang} where it is a very partial implementation for C++, presently goes nowhere near binary object encapsulation for reasons I'll explain shortly. What C++ Modules, as presently planned is as follows:

\begin{itemize}
\item Delay-optimised precompiled library objects are standardised and can be imported at the precompiled binary level (i.e. as GIMPLE for GCC, or as clang-enhanced LLVM for clang) rather than at the uncompiled source inclusion level like \texttt{\#include} requires. These are basically an enhanced form of precompiled headers, plus they are C++ namespace aware unlike the C preprocessor which enforces a completely separate C macro namespace and doesn't understand the code it is working with.
\end{itemize}

And that actually, is it: proposed Modules simply do away with much of the boilerplate made necessary by the C preprocessor header file `hack' used up until now to both export and import library API definitions, though note that you will still have to write header files because they are the source file used to compile a C++ Module interface file which itself is merely rarified auto-generated C++. Note especially that proposed Modules provide no standardised ability to distribute precompiled encapsulated binary objects. Note also that proposed C++ Modules have adopted the same module-local symbol namespace used by the Mach O/ELF approach (which it calls `module partitions'), so while ELF shared objects/Mach O and PE DLLs/COM objects all continue to be strictly speaking necessary, with some extensions one could probably do away with ELF shared objects, potentially even DLLs, \emph{if and only if} the vendor can decide on a stable binary format i.e. it's like asking for precompiled header outputs to work correctly over major compiler revisions, plus be small enough to distribute efficiently.

Note, for later, that a collection of precompiled source code objects (i.e. what C++ 17 Modules proposes) which the compiler can make sense of during compilation and link \textbf{is in fact a dynamic on-disc graph database of AST blobs}\footnote{For correctness' sake, note that Microsoft's compiler is not AST based, though it has been recently gaining AST based parts.}. No one is claiming that that database will be required to be externally usable in C++ 17 -- but I am claiming that compiler vendors will almost certainly have to implement an internal graph database of precompiled module objects for the compiler to examine during source code compilation.

Note also, for later, that a certain amount of link time code generation would probably occur in a world of C++ Modules as it already optionally does in the three main compilers. To what extent one regenerates code during link is where things get very interesting, and to which I will return.

\subsection{WG21 SG7 (Reflection)}

The ability for code to examine other code at the point of compilation entered C++ via template specialisation with the very first standard, and has been constantly improved since then with useful library support in the form of \texttt{type\_traits} and many others. In fact, I would say with some confidence that one of the singular features which separates modern C++ from traditional C++ and all other major object orientated languages is the programmer thinking in terms of \emph{programming the compiler, not programming the program}. Unfortunately the ability for code to examine other code previously compiled (rather than at the point of compilation) has languished somewhat of late, and in the standard that has not advanced past RTTI which lets you only figure out very little about code not known to the current compilation unit.

There have been many proprietary solutions here, with some of the more interesting ones being CERN Reflex \cite{cernreflex}, Boost.Mirror \cite{boostmirror} and Google's Rich Pointers \cite{n3340}, none of which I think it is safe to say cover all the corner cases which can emerge in reflection and for which reason SG7 is initially concentrating on compile-time reflection only (currently I believe they have in mind a C macro-like interface for asking the compiler things about itself which is then exposed via a set of template aliases in a \texttt{std::reflect} namespace, though this situation is still fluid). I'll confess my own involvement here: I tried arguing on the SG7 reflector for having the compiler make available its current internal state into a magic C++ namespace, and then push the problem of dealing with converting that internal compiler state into a reflection API portable across compilers onto third party libraries such as Boost. I basically asserted that designing reflection into the language was too hard, so why not then evolve a library based solution through survival of the fittest? I think it safe to say that this proposal of mine -- which I entitled `Reflection Lite' -- did not see much traction.\footnote{You can read the discussion for yourself at \url{https://groups.google.com/a/isocpp.org/d/msg/reflection/bG52LgeOeGM/NaT1bV4SVaAJ}.}

What I had in mind was easy interoperation with what implementing C++ Modules will already demand from compiler vendors. If to implement C++ Modules the compiler must be able to parse a graph database of precompiled source files at the point of compile, then you \emph{already have the machinery in place to export what is needed to implement a reasonable initial stab at both compile-time and runtime Reflection} -- the only hard part now is deciding how to make available that C++ Module machinery as a Reflection API (my preference) or C++ language feature (probably the consensus preference). I also personally rather like how the exact same solution also works as easily for both compile-time and runtime reflection -- plus, because it's a library based solution, it is very amenable to other programming languages directly interacting with the graph database of modules as-is rather than having to go through bindings to the C++ reflection language feature.

\subsection{WG21 SG8 (Concepts)}

Anyone who has used a heavily metaprogrammed C++ library like some of those in Boost will know well the unpleasantness of figuring out what pages of obtuse compiler error message might mean, plus just how easy it is to cause the compiler to spew such errors for what ought to have been a trivial change. Another big issue with heavily metaprogrammed C++ libraries is compile times -- I remember well a project I worked on in BlackBerry which took seven minutes per compilation unit, and that duration is both just too long and too short to be anything but highly frustrating, trust me. Such brittleness and long feedback loops during debugging makes using heavily metaprogrammed libraries into a very steep learning curve exercise, and quite the tedious chore, for even those highly familiar with the techniques employed. Such are the high fixed initial costs of coming to grips with these libraries that I usually categorically recommend \emph{against} their use to C++ engineers who seek my opinion, and I find that a real shame, especially as apart from these I would rarely advise against using any Boost library.

Original Concepts i.e. the ones proposed for C++ 11 were supposed to let library writers do something about the problems caused by heavy use of metaprogramming mainly by replacing the need for metaprogramming, but they were removed from the 2011 proposed standard as they hadn't been fully thought through. For the 2017 proposed standard, an alternative to Concepts called `Concepts Lite' (its TS is before the committee \cite{n3889}, so some form of it will almost certainly be in C++ 17) has been formulated instead, which has seen much wider acceptance probably due to their novelty being in their lack of novelty.

So what are Concepts as they will enter C++ 17? Well, any substantial changes between now and 2017 notwithstanding, all they do is to let the programmer (and usually actually the library writer) specify, without as much bending over backwards as is normally necessary, \emph{constraints} on template argument types. Everyone is familiar with the use of class and typename typed template arguments:

\begin{program}[!h]
\centering
\caption{Pre-Concepts template arguments}
\begin{lstlisting}[language=c++]
template<typename T, ...> struct Foo ...
template<class T, ...> struct Foo ...
template<template<class, ...> T, ...> struct Foo ...
// And so on ...
\end{lstlisting}
\end{program}

Well, with Concepts you now get new template argument types, with class and typename simply now meaning `unconstrained' and anything not class nor typename meaning `constrained':

\begin{program}[!h]
\centering
\caption{Post-Concepts template arguments}
\begin{lstlisting}[language=c++, morekeywords={addable, comparable}]
template<addable T, ...> struct Foo ...
template<comparable T, ...> struct Foo ...
template<template<addable, comparable, ...> T, ...> struct Foo ...
// And so on ...
\end{lstlisting}
\end{program}

And, of course, you can tell the compiler that an addable type needs to implement \texttt{operator+}, a comparable type needs to implement \texttt{operator==} and so on simply by declaring the appropriate \texttt{constexpr} template predicate functions which return an appropriate boolean. All very straightforward.

So how does this help with compile times? Well, constraints are rather like SFINAE except they are much less work for the compiler: because you provide static rules as to what subset of the graph of available types to consider, the compiler can prune the graph of items being considered very considerably. This radically transforms the many O(log n) or worse searches a compiler must do, and therefore ought to have substantial improvements on compile times for concepts-using C++.

How this could help with metaprogramming brittleness is far less clear, at least to me personally. I can see that in the na{\"i}ve case you will get a much more direct compiler error message when you supply a non-addable type to a template requiring an addable. I don't see, however, that this is much use in a large metaprogrammed library unless the library author is very careful to ensure that the concepts are applied as close to user code as possible such that the concept violations make sense to the library user e.g. by the user facing class and all its APIs being individually implemented by a collection of concept checked implementation classes, thus applying appropriate concept checks to each API use at the point of use, all of which sounds rather cumbersome to write. Even with that, I personally think always-useful compile error reporting isn't feasible at least some of the time, and then users are back to stepping through obtuse library internals to figure out what they've done wrong.

In fact, when I raised this sentiment with Andrew Sutton, who is the main person behind Concepts Lite, in a private correspondence he said this:

\begin{quote}
I have serious doubts that concepts will help with the brittleness of
metaprogramming for the simple fact that metaprogramming is brittle. It
was never the goal of Concepts Lite to improve that aspect of C++ usage. A
major goal is to lessen, if not eliminate, the need for metaprogramming to
express template requirements. We want programmers to spend less time
programming the compiler and more time programming programs (to use your
own words... although that sounds very, very much like something I've
said before).
\end{quote}

Andrew is right that metaprogramming is brittle, and I would assume it will remain so given that you are misusing a template syntax not designed for what you are using it, and \texttt{constexpr} is not much better in also using the wrong kind of syntax for functional programming. However, upon further reflection it occurs to me that it is the brittleness of the \textbf{end use} of metaprogramming which is actually the complexity management and scalability problem in C++ i.e. the library users, not so much the library writers. As much as I would like help with writing metaprogramming, what I actually \emph{need} is for a library like Boost.Spirit to produce useful error messages telling you exactly what is wrong with your grammar, and that alone would absolutely transform the use of metaprogrammed libraries -- rather than metaprogramming itself -- into something which could be categorically recommended to all.

It seems to me that it must be feasible to use concepts with other facilities to make automated generation of metaprogramming tractable, and thus skip the end user ever having to debug pages of metaprogramming induced compiler errors. To take Boost.Spirit again, instead of the user writing a pseudo-EBNF grammar in C++ which is brittle and tedious, have them write real EBNF which is then converted into however many tens of thousands of automatically generated template classes needed, all strung together with concepts. This wouldn't have been feasible in C++ 03 certainly because of compiler limits, and I suspect even in C++ 14 it would internal error the compiler for a grammar of any complexity due to heavy use of SFINAE. With concepts though, much of the complexity imposed on the compiler goes away, and perhaps this much more user friendly way of implementing Boost.Spirit becomes possible.

I guess we shall see, but even if not, there is absolutely no doubt that C++ with concepts is a far better complexity-tamed beast than C++ without concepts. I welcome them unreservedly.

\section{What C++ 17 is leaving well alone until later: Type Export}

I ought to raise at this point the demons made manifest by that infamous 2003 paper N1426 `Why we can't afford export' \cite{n1426} which has been so incredibly influential -- you can see that much of what makes the clang compiler very different in design from all other compilers is a reaction to what N1426 so clearly illustrates as to the can of worms one opens when you allow compilation units to see, at compile time, implementation in other compilation units. One Definition Rule (ODR) violations are prohibited by the C++ standard and are automatic undefined behaviour territory, yet they are widespread in any C++ codebase of any size whether hidden via static storage, anonymous namespaces or DLL symbol visibility boundaries, because the inclusion model of type definition (this is where you use \texttt{\#include} to import externally implemented code) makes it far too easy to allow ODR violations, indeed even to intentionally use them to save time and effort (for example, ELF and DLL shared library binaries naturally follow from the inclusion model). I have many times heard claims that `my code has no ODR violations', to which I answer `have you read every source code line of every library which enters your process?' and if they answer yes, you can then hit them with `well what about libraries your \emph{users} decide to load into a process without telling you?' to which they now normally answer something about that being the user's problem.

I personally think that passes the buck: it should be \textbf{easy} to write code in C++ which works sanely in \emph{any} ODR violating scenario without all the hoop jumping and brittleness currently induced, and in my experience this is a huge part of why large code bases such as the Linux kernel or the Python interpreter continue to refuse to use C++, and stick with C, because looking in from outside it looks like we cannot put our own house in order. By to date ignoring the ODR problems raised by exported templates in C++, it is my opinion that \textbf{we are holding back the conclusion of replacing C with C++}, with all the productivity consequences to the wider software industry therein.

One of the most common cases of unintentional ODR violations in real world large scale C++ codebases is version mismatches in loaded dependencies, with those dependencies often buried so deep down in other dependencies that no one has noticed the mismatch. Let us look at a contrived example:

\begin{program}[!ht]
\centering
\caption{In C++ Module Foo v1.0 which is a dependency of Module A}
\begin{lstlisting}[language=c++, morekeywords={size_t}]
// Interface code
namespace Foo {
  class string; // string is opaque
  extern const string &getString();
  extern void printString(const string &);
}
// Internal implementation
class string
{
  size_t length; // length before storage
  char *storage;
};
\end{lstlisting}
\label{Foo10}
\end{program}

\begin{program}[!ht]
\centering
\caption{In C++ Module Foo v1.1 which is a dependency of Module B}
\begin{lstlisting}[language=c++, morekeywords={size_t}]
// Interface code
namespace Foo {
  class string; // string is opaque
  extern const string &getString();
  extern void printString(const string &);
}
// Internal implementation
class string
{
  char *storage; // storage before length
  size_t length;
};
\end{lstlisting}
\end{program}

\begin{program}[!ht]
\centering
\caption{In Module A}
\begin{lstlisting}[language=c++, morekeywords={import}]
// Interface code
import Foo; // the v1.0 Foo

namespace A {
  extern const Foo::string &what();
}
\end{lstlisting}
\end{program}

\begin{program}[!ht]
\centering
\caption{In Module B}
\begin{lstlisting}[language=c++, morekeywords={import}]
// Interface code
import Foo; // the v1.1 Foo

namespace B {
  extern void log(const Foo::string &);
}
\end{lstlisting}
\end{program}

\begin{program}[!ht]
\centering
\caption{In C++ program which imports Modules A and B}
\begin{lstlisting}[language=c++, morekeywords={import}]
import A, B;

B::log(A::what());
\end{lstlisting}
\label{logwhat}
\end{program}

You probably will look at this contrived example and think it could not possibly happen much in the real world -- well, let me give you a real world example which isn't under an NDA. In the Python programming language you are able to create loadable Python modules which can be written in C, C++ or anything else which can meet the Python module ABI requirements. Boost.Python is a very useful C++ framework to save Python module writers a great deal of work when writing in C++, but imagine to yourself what happens if a Python program loads module A which was linked against Boost.Python v2.0 and module B which was linked against Boost.Python v2.1? What will happen now is one of these three outcomes: (i) the interpreter will segfault at the point of loading module B (ii) you'll successfully import both modules okay and \emph{two} slightly different versions of the Boost.Python DLL will enter the Python interpreter's process space with two sets of runtime state (iii) \emph{the first} Boost.Python DLL loaded will enter the process BUT the second module will have a different understanding of that runtime's layout than what is actually the case, but not enough to segfault during load.

So far so good, even if a bit of memory corruption may have happened for outcome (iii) above. But imagine things from the perspective of the na{\"i}ve Python programmer. He sees that \texttt{A.what()} returns a \texttt{Foo.string}, and quite reasonably thinks that can be passed as-is into \texttt{B.log()} which also takes a \texttt{Foo.string}. What will happen now is one of another three outcomes: (i) one copy of Boost.Python will refuse to recognise objects coming from the second copy of Boost.Python because they don't share runtimes, leading to some confusion for the Python programmer (ii) the interpreter segfaults from overwhelming state corruption due to mismatch of runtime data layouts (iii) the operation appears to work, but actually has corrupted memory just a little bit further.

One might begin to see now why Python uses C instead of C++ to implement its interpreter, just as most other interpreted languages do, despite that writing C++ is vastly more efficient. The problem is the \emph{maintenance costs} in environments allowing third party loadable modules compiled by that third party, and this is a cost where C beats C++ hands down.

Of course there are many mitigation strategies here, everything from library versioning through to better documentation, and of course better design will always solve more corner cases. But that isn't the point: the point is that it is very, very easy for multiple versions of -- let's say the Boost library -- to slip unnoticed into very large C++ code bases. In fact, the Boost C++ Libraries are a \textbf{particularly} good example of this problem because most of the libraries are header-only, so a team can bundle in some version of Boost and \emph{no one can tell that they have simply by examining the exported API/ABI}. As different teams cycle the versions of their particular Boost at different speeds, voil{\`a}, instant unknown and unknowable ODR violation.

This problem with coping well with resolving ODR violation is what sank the now deprecated `exported template' feature of C++ 98, and upon whose difficulty of implementation in the Comeau/EDG compiler N1426 expounds in depth. Export required the compiler to implement a new type instantiation model which N1426 calls the `export model' whereby clusters of interacting (template) types were assigned into compilation units which had nothing to do with traditional source code based compilation units, and then a set of complex and occasionally contradictory rules translated between two quite dimorphic type instantiation models, such that what the programmer intended (the inclusion model) was emulated by the export model. Some interesting consequences were that in order to save on disc space, every time you changed the \emph{use} of a template type, all uses and the original implementation of that template type had to be recompiled, which of course was exactly opposite to the original mandate for having exported templates in the language at all. It was therefore rather slow.

So we dropped Export, and made that whole can of worms go away. However from the perspective of non-C++ users -- for example the Python interpreter above -- \emph{all} of C++ looks just as exported templates did to the rest of C++: one giant failure to deal with ODR violation e.g. by replacing the need for the one definition rule, or indeed the inclusion model, with something less primitive.

That, however, was over a decade ago, and we now have clang, whose design is far better suited for implementing exported templates. The top of the range CPU at the time of implementing Export is considerably slower than a first generation Intel Atom -- and slower than the CPU in your smart phone. We easily have sixteen times the RAM, twenty times the storage and low latency SSDs in a PC, even in a smartphone we have ten times the RAM. In order words, we can probably not worry about storage consumption anymore, and therefore relax the need for clustering type instantiation in an Export implementation.

Not -- let me be clear -- that I actually think that originally proposed exported templates is worth implementing. For one thing, the world has moved a lot onwards since; secondly the feature wasn't and still isn't particularly valuable in practice, because as I mentioned earlier in the section on ELF shared objects, having a global symbol namespace always creates more problems than it solves, and besides the C++ Module partitions mentioned earlier are a clear interface contract with module users: type export must stop at the partition, which rather eliminates their purpose. Moreover, why would you bother trying to pseudo-hide template implementation anyway?

However, there is a far more valuable feature closely related to the issues surrounding exported templates: \textbf{code change ripple management}. This is because if you change type \texttt{Foo} in some source file, the ripple effects of the consequences of that change can be \emph{calculated} if you had a full ODR violation resolution machinery in place, noting for later that the direction of such a machinery acts exactly opposite to the machinery which EDG employed to implement exported templates. And that potentially enables some very exciting features, to which I shall return in the next section.

For now though, no one is currently planning to substantially deal with the wider questions nor the wider issues surrounding exported templates for the 2017 standard, despite their dire impact on external users considering the choice of C++ over other language options. We all know why we can't afford Export, but if you look closely, we have started to subconsciously avoid even going near the problems raised by Export when thinking about the future of C++, which I will call \textbf{`the hidden Export problem'}. This phenomenon is in my opinion ever more frequently raising itself as showstoppers for all sorts of things we could really do with in C++, most especially improved tools, methods and processes for dealing with C++ complexity. They are, in my opinion, particularly getting in the way of Modules, Reflection and Concepts, and the 2017 standard will have in my opinion impoverished implementations of all three features because we don't want to go further until we have good solutions for the problems raised by N1426 over a decade ago.

\subsection{What ought to be done (in my opinion) about this `hidden Export problem'}

Note that the sledgehammer of versioning every lump of code such that Foo v1.1 claims ABI compatibility with Foo v1.2 is certainly one way of doing this, despite the well-known `DLL hell' brittleness which results. However I think we can do a lot better, especially as C++ Modules needs to dump the AST blobs onto disc anyway, so we really ought to leverage those AST blobs properly.

And here is where things start to become exciting. If you bring together everything I have talked about in this paper so far: how hardware growth is going to become linear, how that implies that there will be a big return to systems programming languages to eke out more performance, how C++ 17 will hopefully gain Modules, Reflection and Concepts, how Type Export and ODR violation is so known to be \emph{verboten} we have started mentally skipping over their related issues without realising, and how that laziness appears to external users considering the use of C++ over other options as we can't put our own house in order, some awfully interesting thoughts begin to circulate.

For example, \textbf{what if all C++ were compiled as if every bit of C++ were header only?}

Let's go further: what if all C++ were compiled as if \emph{all} code which could interact with that C++ were header only? That includes Python, PHP, Lua, C\#, whatever.

Let's go even further: what if all C++ were compiled as if all code \emph{including that in related processes} which could interact with that C++ were header only? As in, including processes which talk to your process via some form of IPC, such that the system could correctly figure out that all intermediate code correctly adapted to your change, over the IPC transport and serialisation connection, and only when into the other process' code it realises that your change would break something?

Think of the advantages. It would mean that \emph{at the point of compile} you would get an error message from the compiler if your change broke some bit of Python code running in a process which talks to your process via a RESTful HTTP interface. That's right: as you write code, the compiler can calculate -- in a finite time -- the consequences of your changes on \emph{every single bit of code your change touches}. In one fell swoop, unintentional ABI breakage goes away forever -- and C++ would finally have a sane ABI management solution, and moreover one which very strongly recommends it to all other programming languages \emph{because C++ has just become the foremost solution for managing change ripples in large code bases}.

Now, I'll be honest here -- this idea of making a post-2017 C++ treat all code as header-only came about during extensive email discussions regarding an early draft of this paper with Andrew Sutton and Stephan T. Lavavej in quite literally the five days preceding this conferences, thus sparking a very hasty refactoring of the talk and this paper -- in fact, it was Stephan who articulated it first almost as a semi-serious \emph{reductio ad absurdum} distillation of his opinion on my early draft. As the days went by and the idea bounced around my head, it began to occur to me that that was in fact \emph{exactly} what I was driving at in trying to explain why an embedded graph database needs to be done now in preparation for the future.

What you would need to do to implement `as if everything header only' is to break up compilation into per-type units, so let's say I declare a type \texttt{template<class T> struct Foo}. Perhaps some parts of \texttt{Foo} are type invariant, so you might output a bit of AST and clang-LLVM for those, perhaps even as a mini-C++ Module. Then for every use of \texttt{Foo<T>} you would output another AST and implementation in clang-LLVM, dumping all of these little bits of precompiled implementation and ASTs into a graph database with vertices connecting up their relationships, perhaps also as fully formed mini C++ Modules where possible.

I would imagine that this way of compiling C++ is very similar to how GLSL shaders are compiled on demand into the right opcodes for your local video card, and I would point out that such a task is very amenable to being assisted by a cloud compute resource or a CI server -- it certainly is embarrassingly parallel.

Think of it this way: Facebook provides a `Social Graph' which connects up the relationships between billions of individuals. You're doing exactly the same thing here, but with modules of implementation of C++ types, so the node representing the definition of \texttt{template<class T> struct Foo} is connected to the nodes representing all partial and concrete instantiations of \texttt{Foo} such that the consequences of a change to one part can be calculated on the wider graph. Incidentally, a C++ compiler already does most of this as part of converting a compiland's AST into optimised output, just it's an in-memory database rather than being stored to disc. I don't underestimate the refactoring which could be involved here -- right now compilers can perform the graph reductions part of the optimisation passes on an in-memory database, whereas what I propose probably requires doing so with an on-disc database instead, which probably means a lot of new code\footnote{As a v1.0 implementation though, we could probably accept unoptimised code for this feature i.e. it only works when writing code being compiled as debug.}.

The hard part of course is designing the ODR violation resolution machinery, just as it was for Export. We have some advantages over Export though: firstly, we aren't trying to adhere to a design written in stone before anyone had tried an implementation. Secondly, we have a ton more existing complex C++ codebases out there to test against, so instead of having to worry about what someone might potentially do with the standard (remember Export was being implemented before the C++ 03 standard), we now have a fairly good idea. Thirdly, Export defined an additional method of making template types visible outside their implementation which meant that things like Koenig lookup had to start traversing compilands, whereas here we aren't doing that: the inclusion model is always respected, so if you define twenty incompatible implementations of type \texttt{Foo} in twenty separate source files, and then try to use multiple implementations at a single point, we will quite rightly refuse to compile instead of trying to jury rig Koenig lookup to work. Fourthly, we now have a working definition of C++ Modules to hand, and while I can see some potential incommensurability problems with making every fragment of type instantiation its own C++ Module, I don't think they are insurmountable.

I appreciate that that last paragraph rather glides over important detail, but I am going to bail out early -- besides, I have no idea if anyone else likes this proposition (I am self employed and live in a country far removed from the centres of software excellence, so who knows what the compiler vendors think). I will say a little more on how to go about managing the ripples of change though, because sometimes you don't want them to simply propagate unhindered.

It occurs to me that how you would implement change ripple management is through `Concept checks for the ABI', so the embedded graph database I am proposing is not only the repository where you store the little fragments of object file outputs, but also where you articulate the rules governing what to do with code ripples. Do you propagate them onwards, figuring out what code needs to be recompiled until the full extent of the consequences of a single code change is made manifest? Or does a preconfigured fixed ABI bounce incompatible changes back to the programmer -- for example, Microsoft or BlackBerry provide SDKs to end users, and for those you might store each SDK release as a fixed ABI, thus ensuring the programmer cannot accidentally break a SDK during a bug fix?

The possibilities are vast, but given the newness of the idea to me, I am unfortunately going to stop with the above description. Perhaps I can expand at next year's C++ Now. What I will say though is that a feasible implementation implies that you can't do a full recompile of all affected code at the point of compile. What we really need here is two things: (i) reduce the search space and (ii) the ability to push the hard graft onto a batch pass, probably done by some cloud resource somewhere, which periodically runs over the binary AST blobs spat out by each compilation unit, and generates indices of things to look for during individual compiles. The compiler or linker can then massively reduce the amount of work that checking for the consequences of changes would require, thus making my proposal tractable. The runtime linker might also coalesce regularly non-changing binary outputs into ELF or DLL shared libraries, and thus avoid duplication of code in each process -- something greatly helped by the fact that the proposed graph database keeps its first tier content as an ordinary file which can be mmapped etc. like any other.

Once again, we are looking at the need for a \textbf{graph database} which can store this sort of metadata, and remember that you're going to have to push a certain amount of this database with programs which need to load third party supplied extension modules, so this is a distribution issue as well, something I've accounted for in the proposed graph database as it can shard bits of itself over a network transport.

Coming back to the Boost.Python example mentioned in the previous section, Boost.Python is very much limited in flexibility by the lack of Reflection in C++ -- basically you must laboriously tell Boost.Python how to make each C++ object available to Python, or more likely, you use a bindings generating tool such as Pyste which statically implements Reflection through parsing your C++ interface files using GCC-XML. If Boost.Python could have the compiler do the static Reflection -- or even more exciting, do \emph{dynamic} Reflection of some arbitrary precompiled C++ Module like you can in Managed .NET C++ -- it would hugely ease the complexity of implementation, the difficulty of generating high quality bindings, and generally substantially improve C++'s usefulness to other programming languages, and finally persuade them to leave C.

If it isn't obvious by now, the very same \textbf{graph database} sitting at the very core of the C++ runtime can provide your repository of linkable and dynamically loadable modules, the ASTs and debug infos which were generated during compilation for Reflection and Concepts synthesis, and the timestamping/versioning of all the previous so you can easily figure out what relates to what and when.

All this also goes a huge way to letting build systems become far more optimal than at present, as instead of a make tool recompiling all dependencies of a changed file, we could rebuild on the basis of actual consequence of a change i.e. if you change something which has no consequence on external type dependencies, nothing gets rebuilt. This ought to let C++ scale up nicely into ever larger code bases for at least another decade, despite the linear growth of CPU and RAM capacities.

\section{What I am pitching: An embedded graph database at the core of the C++ runtime}

Back when I first began to ponder solving information complexity management at an industry-wide scale in the 1990s (and ended up unknowingly reinventing Plan 9 to solve it\footnote{\url{http://www.tnrev.org/}}), I placed at the very heart of my next-generation platform design a shard of a distributed global active object database, rather similarly to how Plan 9 did by making everything in the distributed system a file. Everything -- every file potentially loadable into a program, every possible interpretation of that file (e.g. text as paragraphs), view and conversion (e.g. PNG as a JPEG) of every possible piece of data, and the history of what you did as well as everyone else did (which was part of the capability token-based security model) -- all appeared in that object database shard which was lazily evaluated, so diving into some subdirectory was equivalent to constructing a lazy demand-executed operation. I ended up making only a bite into that implementation in the three full time years I invested into it, but I have been very glad to watch `the Web 2.0 cloud' manifest most of that solution before all our eyes. What surprises me continuously however, is how the internet of graph connected information, which anyone can so easily tap into with a RESTful HTTP API request, \emph{is absolutely absent from local devices}, even to the software developer. The closest thing I can think of which approximates an ubiquitous local database is probably SQLite3, which while a superb piece of software I have made much use of over the past decade and a bit, is not a graph database, even with an ORM translation layer mounted atop.

I find this situation to both be rather curious as well as a bit of a shame: for example, if C++'s shared library loader could construct a graph of load ordering dependencies, it could load independent graphs simultaneously using multiple CPU cores. As the serialised nature of process initialisation is a very significant limiter of process initialisation speeds in large C++ codebases with many shared libraries, this could make feasible \emph{micro-}shared libraries -- as in, consisting of one or two very small, very specialised and very reusable class implementations somewhat resembling the instantiation clusters of EDG's implementation of Export -- with future C++ applications loading thousands, or \emph{tens} of thousands of these tiny shared libraries into a process space. Obviously the granularity of the system page size (usually 4KB) is a factor here, but one could see that an `as if everything header only' implementation could simply spit out lots of little shared libraries, and let a runtime linker figure out which subset is the right subset for the current process.

Anyway, if having established how useful an embedded graph database in C++ would be, you then proceed to look around for an embedded graph database implementation, you will be perhaps surprised to hear that I could find precisely one: UnQLite\footnote{\url{http://unqlite.org/}}, which happens to come from the same stable as SQLite, and it is one very impressive and useful piece of software. However, in my opinion I don't think it can substitute in the same wide ranging use cases as SQLite, and here are my reasons why:

\begin{enumerate}

\begin{table*}[!ht]
\centering
\begin{threeparttable}
\newcommand{\LosesData}{\color{red}{Yes}}
\newcommand{\DataSafe}{\color{green}{No}}
\caption{What non-trivially reconstructible data can you lose if power is suddenly lost for various popular filing systems?\tnote{1}}
\begin{tabular}{R{2.3cm}|C{1.6cm}|C{1.6cm}|C{1.2cm}|C{1.5cm}|C{1.8cm}|C{1.5cm}|C{2.7cm}}
& Newly created file content corruptable after close\tnote{2} & File data content rewrite corruptable after close\tnote{2} & Cosmic ray bitrot corruptable & Can `punch holes' into physical storage of files\tnote{3} & Default maximum seconds of reordering writes of newly created files & Default maximum\tnote{4} seconds of writes loseable & Measures necessary to safeguard data even after waiting max seconds of writes loseable\\
\hline
FAT32                                  & \LosesData & \LosesData & \LosesData & \color{red}{No} &  ? &  ? & (i) Parity data \newline (ii) fsync dir \newline (iii) fsync data \newline (iv) vanishing/duplicate file entries check\\
\hline
ext2                                   & \LosesData & \LosesData & \LosesData & \color{red}{No} & 30\tnote{5} & 35\tnote{5} & As above\\
\hline
ext3/4 \texttt{data=writeback}         & \LosesData & \LosesData & \LosesData & \color{green}{ext4 only} & 30\tnote{5} & 35\tnote{5} & As above\\
\hline
ext3/4 \texttt{data=ordered} (default) & \DataSafe  & \LosesData & \LosesData & \color{green}{ext4 only} & 30\tnote{5} & 35\tnote{5} & (i) Parity data \newline (ii) fsync rewrites \\
\hline
UFS + soft updates                     & \DataSafe  & \LosesData & \LosesData & \color{red}{No}\tnote{6} & 30 & 30 & (i) Parity data \newline (ii) fsync rewrites \\
\hline
HFS+                                   & \DataSafe  & \LosesData & \LosesData & \color{green}{Yes} & ? & ? & ? \\
\hline
NTFS                                   & \DataSafe  & \LosesData & \LosesData & \color{green}{Yes} & idle/\newline write limit & idle/\newline write limit & (i) Parity data \newline (ii) fsync rewrites \\
\hline
ext3/4 \texttt{data=journal}           & \DataSafe  & \DataSafe  & \LosesData & \color{green}{ext4 only} &  5\tnote{5} &  5\tnote{5} & (i) Parity data \\
\hline
BTRFS                                  & \DataSafe  & \DataSafe  & \DataSafe & \color{green}{Yes} & 30 & 30 & \\
\hline
ZFS                                    & \DataSafe  & \DataSafe  & \DataSafe & \color{green}{compress only} &  5\tnote{7} & 30 & \\
\hline
\end{tabular}
\begin{tablenotes}
\item [1] I should stress that this table has been constructed by me through a lot of `best guesses' and examining implementation source code where available. I am not at all confident in its accuracy. Note the complete lack of information about Apple's HFS+ filing system, other than its metadata being journaled there is very little information out there. I also could not find much which is concrete about Microsoft's NTFS filing system -- we know it will pace writes after RAM buffers exceed some amount, and we know it will early flush RAM buffers if the device is considered idle, past that I couldn't find out much more.
\item [2] `closed' is a very important qualification: only a few filing systems guarantee (rather than `just happens') anything about writes to files without all file handles closed.
\item [3] This is where a filing system permits you to deallocate the physical storage of a region of a file, so a file claiming to occupy 8Mb could be reduced to 1Mb of actual storage consumption. This may sound like sparse file support, but transparent compression support also counts as it would reduce a region written with all zeros to nearly zero physical storage.
\item [4] This is the maximum time before the system will start to try writing dirty data out to storage. It may start to write the data sooner e.g. if a large amount of dirty data has not yet been written. It also may take a considerable period of time before the dirty data actually reaches storage.
\item [5] The ext2/3/4 filing systems commit their metadata by default every five seconds, however ext3/4 will not commit metadata for file data not yet allocated and ext4 may significantly delay allocating storage. This, in practice, means that ext3 may take \texttt{dirty\_writeback\_centiseconds} (default=5s) to start writing out dirty pages, whilst ext4 may take \texttt{dirty\_expire\_centisecs} (default=30s) before allocating space for written file data, which then gets written out the same as ext3.
\item [6] UFS itself supports sparse files, yet I could not find an API with which you can punch holes for arbitrary regions.
\item [7] The chances are that the ZFS write throttle is going to get replaced once again, so the fixed five second per transaction group which limits write reordering to five second chunks is going away. See \url{http://dtrace.org/blogs/ahl/2014/02/10/the-openzfs-write-throttle/}.
\end{tablenotes}
\label{powerlossmatrix}
\end{threeparttable}
\end{table*}

\item UnQLite does too much: it comes with a custom scripting language which is compiled into a bytecode, pluggable storage engines, ACID transactions, a HTTP request engine, plus some funky C idioms to extend C with desirable C++ features. I have no problem with any of that, and it makes substituting UnQLite for a big iron commercial graph database far easier, but if there is anything which I have learned from writing open source libraries, it is that \emph{people far prefer small, flexible, reusable \textbf{bits} of implementation rather than anything approaching a \textbf{complete} implementation}. My most popular open source code by far is nedtries followed distantly by nedmalloc, which between them make up only hundreds of lines of code compared to the 50k+ lines of open source of mine out there.
\item UnQLite enforces the key-value model on you. There is nothing wrong with that, it is very popular, but I suspect that many developers will want to use whatever arbitrary custom graph layout they feel like -- or even more likely, to end up \emph{evolving} some graph layout which fits their needs over time. My point is that an embedded graph database really shouldn't impose anything like a data model on you because it's so low level: you should be free to model and index as you want.
\item UnQLite stores its data in a single file in a custom format. I suspect that many developers will prefer in an embedded graph database to directly open a file containing the right data on the hard drive, and really all a read-only embedded graph database should do is to tell you what the path of that file is.
\item UnQLite doesn't appear to natively version your data for you such that you can rewind history, nor let you easily recover from bit errors caused by cosmic ray bitrot (something surely very valuable in an embedded database!). Closely tied to lack of versioning is the lack of secure modification auditing whereby any attempt to change history will rewrite all subsequent history, something very useful for a multitude of uses.
\item UnQLite doesn't easily merge shards of databases over unreliable links (or if it does, I didn't find such facility). I'm thinking something like `git push' and `git pull' would be very useful for sending bits of graph database around the place, copying only those parts not in the destination database.
\item UnQLite's design doesn't take advantage of useful filing system specific features -- for example, consider the power loss safety matrix in Table \ref{powerlossmatrix} which shows what non-trivially reconstructible data can you lose if power is suddenly lost for various filing systems. With a proper design, if your graph database were residing on a ZFS volume which makes very strong guarantees about data integrity and write ordering, then you could safely \emph{completely skip \texttt{fsync()} without risking data integrity of some past complete history of versioned data} and only risking the last thirty to maybe sixty seconds of data written. If ZFS sounds a bit exotic, consider that you can eliminate \texttt{fsync()} completely on an ext4 partition mounted with \texttt{data=journal} if you are willing to risk losing the last five and a bit seconds of data written, and do bear in mind that one can absolutely keep your graph database in a specially mounted ext4 partition.
\item Like most key-value document stores, UnQLite does let you make database objects active via its scripting language, but you are limited by the features of that scripting language -- which is of course not C++.
\end{enumerate}

So here is what I am thinking instead: why not reuse the same content-addressable object storage algorithm as used by git to implement a very low-level transaction-capable data persistence library, and then allow users to bolt some arbitrary per-graph or per-tree indexing implementation on top to implement the lookup indices (one of course includes Boost.Graph here, but one could certainly include a git implementation and SQL table implementation too)? Such a graph database would not be particularly quick as it would be entirely based on files in directories on the filing system, but it would be extremely low-level and I should imagine could still achieve no less than 50 transaction commits per second on a magnetic hard drive which ought to be plenty for many applications. What users do with such a low-level graph database is entirely up to them: don't want transactions? No problem. Don't want Boost.Graph? Also not a problem. Don't even want versioning? Well, we can do even that too.

\pagebreak

\subsection{A quick overview of the git content-addressable storage algorithm}

The git source control tool has become known to most software developers, and of course this year Boost moved onto git which was a huge achievement. What is less familiar to most users of git is how cleverly stupid the git implementation is (which is why Linus named it `git' incidentally, a git is an annoying stupid person in British English), so a quick overview is in order.

Most know that git works exclusively with SHAs, these being a 160 bit long number which are the SHA-1 of the content in question, with the assumption that these never clash and so that number is a unique identifier of the content in question. Content is then stored inside the git repo under \texttt{.git/objects/ab/xxxx} where ab is the first two letters of its SHA in hexadecimal, and xxxx is the remaining 38 characters of the SHA in hexadecimal.

Directories of files and their metadata such as timestamps are simply made up of a list of leafnames and metadata and the SHAs of the content of those files. These are stored in a structure called a tree object, which itself is SHAed and stored alongside the content as just another blob of content. A directory entry in a tree object of course refers to another tree object, thus you get the standard directory hierarchy.

Commits consist of a linked list of commit objects, where a commit object is simply the SHA of the appropriate tree object, preceding commit object and commit message. Branches, which are simply a reference to some commit, are quite literally a file in \texttt{.git/refs/heads} named after the branch solely containing the SHA of the commit in question.

Using just the information above, you can now write a fully functional git implementation which will be understood by any other git implementation (though other git implementations will almost certainly convert your repo into something you can't understand due to an extra feature called packfiles, but these aren't germane to this discussion).

The especially clever part about git starts to become clear when you start thinking about pushes and pulls and other sorts of atomic concurrent updating by multiple processes. When pushing, all git does is to send a list of SHAs making up the commits of the branch being pushed. The remote git compares the SHAs in the list to see which ones it already has simply by testing for file existence, and it acks with the list of the SHAs it doesn't have, thus only transferring the minimum delta of data necessary. Pulls (really fetch) simply do the same thing in reverse. Note a \emph{very} useful property: if the transfer of missing SHA blobs gets interrupted, you get for free resumable transfers.

This then raises an interesting question: what happens to orphaned SHAs i.e. those which were part of a failed transfer, or orphaned due to a branch being deleted? Well git solves that using garbage collection which it runs periodically: it simply makes a list of all SHAs in all branches, commits and trees still being used, and deletes all those file entries in objects it finds on the filing system not in that list. This is why in git even if you forcibly trash a month's worth of other people's work by doing \texttt{git push -f} as some are too fond of doing, all your content is still in there and very easily retrieved e.g. simply reset the branch HEAD to whatever SHA it was before and yes, it just works so long as no garbage collection has happened since.

Even more clever again in this very stupid design is how git handles concurrent writes by multiple processes. Because content is uniquely named according to contents, if two processes write the same filename you are guaranteed that it is the same data -- all you need to do when writing new content objects is to open using \texttt{O\_EXCL} and you're good. Rather more interestingly of course is that simultaneous writes to dissimilar content generally proceed in parallel so long as no one writes the same content (with reads completely unhindered), and even there you simply push the failure to exclusively open an item to the end of the list to check again to see if it's finished. In fact, the only time you actually need to serialise at all is when writing concurrently to the same branch: there commits must be sequential, so some commit must come first -- even in this case, you can safely let the two commits write their content in parallel and if you later realise you need to recalculate one of the commits, you get partial resumption of that recalculated commit, writing \emph{only} the data recalculated.

\subsection{How the git algorithm applies to the proposed graph database}

You're probably beginning to understand how more or less the same algorithm -- albeit somewhat less hardcoded -- would make for a pretty handy ACID-compliant transactional database. Instead of branches, think named graphs, and instead of storing a single hash in the graph HEAD file, keep a list of hashes in a hole-punched file opened as append-only\footnote{Files opened append-only are guaranteed to be updated atomically across all open file handles on all major operating systems. The hole punching lets you prevent all physical storage from being consumed by a constantly `growing' file. It may yet be an implementation option for reasonably fast transaction locking over networked filing systems.} which brings you versioning. If you want to atomically update more than one graph at once in a single transaction, you can use the same optimistic parallel write scheme with lazy recalculation of ordering induced changes. If you suddenly lost power, you can figure out the last known good state very easily by parsing all the trees referred to by hashes starting from the end of the HEAD file going backwards and finding (on those filing systems requiring it, see Table \ref{powerlossmatrix}) which was the most recent version to have a perfectly uncorrupted state\footnote{A good question remaining is whether a write journal might be superior -- it may well be over networked filing systems.}.

There are, of course, still many unresolved implementation questions: How do you safely prune the version history from the append-only graph HEAD files on filing systems without hole punching support whilst being heavily used by multiple processes without incurring writer starvation? Do you allow users doing i/o on the graphstore using network filing systems such as NFS and Samba, and if so how do you downgrade machine-local processes to use the right NFS and Samba safe locking scheme instead of a shared memory region? Indeed, how do you get Windows and POSIX to lock files in a portable way, as the two schemes are quite incommensurate?

Also, what about all those files being written onto the filing system? What about cluster wastage (this is where a very small file will still consume the cluster size e.g. 4Kb on some filing systems)? What about hitting inode limits (some filing systems `run out' of total files possible on a filing system)? How atomic and power loss-safe are file and directory renames really?

What about performance? Performance won't be in the league of big iron graph databases by any stretch, but I can see 10,000 transactions written per second to different graphs as being very possible on a modern non-Windows\footnote{Microsoft Windows currently has an approx. 30,000 maximum file handle opens per second limit on a quad core machine, which is several thousand times slower than any other operating system.} filing system if you are willing to lose recent writes by turning off fsync. ZFS on inexpensive hardware with a fast write intent log device will let you do 30,000 \emph{synchronous} (as in, fsynced) file creations per second until the intent log device fills up, so somewhere between two and five orders of magnitude slower than the big iron databases might be expected for copy-on-write filing systems whilst retaining perfect data persistence.

To round off this position paper, I'll move onto a non-hand-wavy concrete technical proposal as a demonstration of just one of the many useful things C++ could do with an active graph database in its core runtime: object components.

Useful additional reading about persisting and locking data reliably and portably:

\begin{itemize}
\item \url{https://www.sqlite.org/lockingv3.html}
\item \url{https://www.sqlite.org/wal.html}
\item \url{http://www.westnet.com/~gsmith/content/postgresql/TuningPGWAL.htm}
\item \url{http://www.postgresql.org/docs/9.1/static/runtime-config-wal.html}
\end{itemize}

\section{First potential killer application for a C++ graph database: Object Components in C++}

One of the reasons I have repeatedly dwelt on C++ Modules during the previous discussions is that I believe we can not only do better than Modules, I believe we can \emph{transcend} Modules at a not too hideous a cost to the industry, and I believe that one of the ways of really taming large code base change ripple management in C++ is via \textbf{componentisation} of the C++ binary stack. Let me quote Tony Williams, one of the chief architects of Microsoft's Component Object Model, who was trying to persuade Microsoft in the late 1980s to write code resilient to unpredicted change:

\begin{quote}
Our software represents a major capital resource. We need the ability to maintain and evolve our software without destroying its strength ... we need the ability to replace parts of the cellular structure with new ones. Where new meets old, we need well defined shapes for them to join. It will become increasingly hard to remain competitive if we have to dismember the ... structure into its component cells, and rebuild most of those pieces just in order to put them back together. \cite{williams1990inheritance}
\end{quote}

How much the dominance of Microsoft over software in the 1990s and 2000s had to do with this approach is hard to say, but Microsoft itself thinks it was a key factor\footnote{\url{http://www.microsoft.com/about/technicalrecognition/com-team.aspx}}.

It has been such a long time since software components were the \emph{zeitgeist} in the 1990s that I will probably have to explain, for the benefit of younger readers, what a software component actually is and why it is definitely not a Module nor a DLL/shared object. I should also caveat in the following description that no one entirely agrees on what a software component is, but the following should be close enough:

\begin{itemize}
\item Components declare [compatibility with] some sort of event processing model as part of their interface contract.
\item Components often declare a sort of what we would nowadays call a service orientated architecture (SOA), a bit like a proto-web service.
\item Components should \emph{never} need to be recompiled just because external code using them has changed. In fact, any header files used to interface with them ought to be generatable solely from their precompiled binary blob \emph{(note the implied need for an additional information layer separate from the implementation)}.
\item Components are generally designed with an awareness that they will be put to unforeseen uses to a much greater extent than traditional libraries.
\item Components are therefore \emph{highly} reusable by being \emph{very loosely coupled} with the code using them.
\item Components are usually so highly reusable in fact that languages other than that of their implementation can use them easily.
\item Components (for the purposes of this paper) are NEVER connected by a data transport e.g. a pipe, or a socket. That I would call `distributed components' and might take the form of D-Bus, DCOM, CORBA and so on. The kind of components referred to here are \textbf{always} in-process i.e. executable code implementing a majority of the component is loaded directly into the process space where it is consumed.
\end{itemize}

To younger engineers, these sound vaguely familiar in a `I'm not sure why' kind of way: the reason why is because they are so utterly ubiquitous in proprietary vendor implementations they have faded from conscious consideration. One of the most ubiquitous of all is Microsoft's Component Object Model, better known as COM, which has been recently refreshed into a new enhanced iteration better known as WinRT, with the more modern C++ interface to WinRT being called C++/CX. However Microsoft's implementation is hardly the only ubiquitous component implementation, in fact OS X's component object model -- which in many respects \emph{is} the \emph{raison d'{\^e}tre} behind the features of Objective C extending C -- was originally designed by NeXT as their component object broker architecture in the very early 1990s, and whose design has remained (apart from the addition of threading) pretty much untouched since then for some twenty-five years now (and still going strong).

As to which component objects implementation is the more ubiquitous, I am unsure, but we are talking billions of installations for both. And note that outside of the Android/Linux/GNU ecosystem which remains steadfastly components-free at the systems level, between Microsoft's and Apple's proprietary components implementation a good chunk of the world's systems language written software is componentised. In short, object components have been a \textbf{huge} success everywhere outside of the open source software ecosystem, mainly because \emph{they enable software to scale}, yet they have not and are not currently intended to be standardised in C++.

\subsection{A review of Microsoft COM}

As Microsoft COM is regularly used by C++ and Apple's is not, I shall restrict myself henceforth to discussing Microsoft COM. Microsoft COM was very intentionally designed at the time to NOT use C++ or any feature of C++ \cite{box1998essential, williams1990inheritance} because taming the C++ complexity (of the 1990s) was seen as the primary goal. Influenced strongly by DEC VMS's interface definition language IDL (the next-gen DEC VMS kernel became the NT kernel), object inheritance was split into two categories: (i) implementation inheritance and (ii) interface inheritance, with COM intentionally only implementing the latter kind and specifically not the former. A component's interface could be written up in IDL (which in practice is auto-generated for you by the toolset), that IDL compiled into a binary interface description and linked into a standard DLL and voil{\`a}, you have yourself a COM object.

To illustrate how COM works in terms of code, consider that the C++ code in Program \ref{comcppcode} would `expand into' an effectively identical C code in Program \ref{comccode} when compiled, this being due to the fact that a good chunk of C++ of that era was still being compiled into C by a tool called `CFront'.

\begin{program}
\centering
\caption{Example C++ code}
\begin{lstlisting}[language=c++, morekeywords={}]
class Foo {
  int a;
  virtual int func() { return a; }
};

int main(void) {
  Foo n;
  return 0;
}
\end{lstlisting}
\label{comcppcode}
\end{program}

\begin{program}
\centering
\caption{Equivalent to Program \ref{comcppcode} in pseudo-C}
\begin{lstlisting}[language=c++, morekeywords={}]
struct Foo;
struct __vtable_Foo_s {
  const size_t base_class_offset;
  const type_info *rtti;
  const void (*byIndex)(Foo *)[1];
};
int Foo_func(Foo *this) { return this->a; }
/* We'll leave out the type_info init for brevity */
static const type_info Foo_RTTI;
static const struct __vtable_Foo_s __vtable_Foo={0, &Foo_RTTI, &Foo_func};

struct Foo {
  const struct __vtable_Foo_s *__vfptr[1];
  int a;
};

int main(void) {
  struct Foo n={&__vtable_Foo};
  return 0;
}
\end{lstlisting}
\label{comccode}
\end{program}

Microsoft COM adopted the C++ vtable layout as the same for COM, so C++ classes could be easily wrapped up as COM objects i.e. the pointer to the statically stored read-only \texttt{struct \_\_vtable\_Foo\_s} in Program \ref{comccode} is directly exported as-is. Note as made clear above that the C++ class vtable is simply a C struct of function pointers, so C code or anything which can speak C code has no problem calling into the COM object implementation, in fact that is exactly what the COM IDL outputs for COM C bindings.

Microsoft COM is, in fact, remarkably simple considering how hideously difficult it was considered to be back in the 1990s (and vital to master if one wanted to preserve one's career back then). The \textbf{only} thing COM exports from a C++ class is the class vtable i.e. its list of virtual functions. \emph{And that's it} -- no data, no non-virtual functions, certainly nothing resembling a template nor is any form of structured exception passing supported\footnote{C++/CX, a.k.a. `WinRT C++' has mangled in a way of pushing exceptions across a COM boundary, but COM itself still returns a simple C style integer error code.}. Rather more germane to the wider point I am making, \emph{you are not allowed anything from C++ which is compile-time} i.e. most of the additions to C++ since the 1990s, so if you'd like to pass a \texttt{std::tuple<...>} as a parameter to a C++/CX object member function without adding a vtable and reference counting to \texttt{std::tuple<..>} (i.e. removing its compile-time nature by forcing it into a concrete implementation, which kinda defeats much of its purpose), you're out of luck.

There are two more items I should mention about Microsoft COM before moving on: the first is the COM event model, mainly because it is a good example of what needs to be avoided in any future C++ components implementation. Reading the Microsoft white papers and books on COM of the time, it is clear that no one deeply considered the `other half' of COM which is its event processing model until quite some years later when serious problems such as the inability to avoid threading deadlocks began to emerge. It didn't also help that COM had grafted onto it after its design a threading model, a distributed execution model, and then of course all of .NET which still had to remain COM-compatible.

One of the most major enhancements of COM implemented by WinRT is a proper, well designed, asynchronous event model which uses a system-determined mix of kernel threads and fibres (i.e. cooperative rather than pre-emptive task switching) to rid multithreaded programming of COM of most of its historical weirdnesses and quirks for newly written code (for which this author is immensely grateful to Microsoft). Some would even say that WinRT is in many ways mostly a substantial upgrade of the very same COM which has been powering the Microsoft software ecosystem for two decades now.

The second thing I should quickly mention is how COM implements its registration database: each COM object gets assigned at least one randomised UUID which is a 128-bit number to uniquely identify it (further UUIDs indicate version). This number is registered with the list of available COM components by writing that UUID along with other information into various special locations within the Windows registry which is a \emph{non-referential} key-value database. Even though the Windows registry implementation has become quite resilient to damage in the past decade, the fundamental lack of referential integrity is a leading cause of Windows installations becoming unreliable over time \cite{ganapathi2004pcs}. This is another good example of what needs to be avoided in any future C++ components implementation.

\subsection{An example design for modern C++ component objects}

\begin{program}[!b]
\centering
\caption{How we mark up a C++ class as a component (backwards compatible)}
\begin{lstlisting}[language=c++, morekeywords={}]
// Type being componentised
namespace Boo {
  class Foo {
    int a;
    Foo(int _a) : a(_a) { }
    virtual int func() { return a; }
  };
}

// The component export metaprogramming declared at interface level e.g. via a EXPORT_LEGACY_COMPONENT(EXPORT_COMPONENT_NAMESPACE(Boo, Foo), ::Boo::Foo)
namespace __component_export__ {
  namespace Boo {
    class Foo : public
  __component_export_machinery__::exported_component<::Boo::Foo, false> {
      static const char * __location_file() { return __FILE__; }
      static size_t __location_lineno() { return __LINE__; }
      __component_export_machinery__::exported_component<::Boo::Foo, false> operator=(const Foo &);
    };
  }
}
\end{lstlisting}
\label{componentmarkup1}
\end{program}

\emph{[This section was written before the `as if everything header only' concept was reached, and I didn't have the time before the conference to refactor it. Consider it for historical interest only, but I have added a set of quick notes at the end]}

I don't claim in the following example to have thought through everything, nor does the following example design support as much as I believe would be possible once we have C++ Modules. In other words, the following example is intended to be a purely illustrative solution showing how object components might work in a C++ graph database, and it in fact is quite similar to John Bandela's CppComponents framework (\url{https://github.com/jbandela/cppcomponents}) except that I don't require a vtable as Microsoft COM and CppComponents does, plus his only implements the ABI part of components with optional COM-compatible memory management, and no event model.

I should also make an important point clarifying Intellectual Property very clear before I begin: I delivered as part of my employment a functioning demonstration prototype implementing some of this example design whilst I was working for BlackBerry 2012-2013, but the design, and a majority of the code making up that demonstration prototype, actually came from my next-generation platform which I mentioned earlier. I was given exactly one calendar month to deliver that demonstration prototype which weighed in at some 12,000 lines of code, and to deliver such a large project in such a short time I had to reuse a lot of old code written back in 2002-2004, which actually turned largely into an interesting exercise in porting legacy code to C++ 11 and Boost. Given the balance between old and new works involved here strongly favouring the pre-BlackBerry IP, I should consider everything I am about to say as originating solely from my next-generation platform, and having nothing to do with my work at BlackBerry -- nevertheless, I should consider the following example design potentially tainted, which is why it is an example design and not a proposed design -- any actual implementation ought to utilise a different design to avoid any possible IP problems later.

The first part is a little bit of metaprogramming to have the compiler tag classes as components to the link stage with metadata only known to the compiler, and it takes the form as shown in Program \ref{componentmarkup1}.

You might be surprised to learn that that is quite literally it for the vast majority of C++ classes -- as is obvious, you don't even need to recompile a library as all you normally need is the header file. The reason we use \texttt{operator=()} is because the compiler will automatically spit out a default implementation anyway, so you might as well halve your symbol counts by reusing it, though note the covariant return type of its base class which is a useful parsing shortcut for later.

The markup in Program \ref{componentmarkup1} is actually the convenience version intended purely for retrofitting legacy codebases. New code really ought to use the markup in Program \ref{componentmarkup2} instead which requires componentised classes to be placed into some internal namespace, then the act of marking up the class as a component will define the \emph{componentised} version of the class for use by the rest of the component's code. There are very significant advantages to doing it this way round instead.

\begin{program}[t]
\centering
\caption{How we mark up a C++ class as a component (non-legacy)}
\begin{lstlisting}[language=c++, morekeywords={}]
// Type being componentised
namespace internal {
  namespace Boo {
    class Foo {
      int a;
      Foo(int _a) : a(_a) { }
      virtual int func() { return a; }
    };
  }
}

// The component export metaprogramming declared at interface level e.g. via a EXPORT_COMPONENT(EXPORT_COMPONENT_NAMESPACE(Boo, Foo), internal::Boo::Foo)
namespace __component_export__ {
  namespace Boo {
    class Foo : public
  __component_export_machinery__::exported_component<internal::Boo::Foo, true> {
      static const char * __location_file() { return __FILE__; }
      static size_t __location_lineno() { return __LINE__; }
      static __component_export_machinery__::exported_component<internal::Boo::Foo, true> __signature();
    public:
      using internal::Boo::Foo::Foo;
    };
  }
}
namespace Boo {
  typedef __component_export__::Boo::Foo Foo;
}
\end{lstlisting}
\label{componentmarkup2}
\end{program}

You're probably now wondering what magic lives in \texttt{\_\_component\_export\_machinery\_\_::exported\_component}, well it's nothing special, and you can see a condensed version in Program \ref{componentexport}.

\begin{program}[!t]
\centering
\caption{Metadata stored by \texttt{exported\_component}}
\begin{lstlisting}[language=c++, morekeywords={}]
namespace __component_export_machinery__ {
  template<typename T, bool resilient, size_t data_size=sizeof(T), size_t vptrs_size=detail::sizeof_vptrs<T>::value> class exported_component {
    virtual void __end_of_vtable() {}
  };
  template<typename T, bool resilent, size_t data_size> class exported_component<T, resilient, sizeof(T), 0> {
  };
  template<typename T, size_t data_size, size_t vptrs_size> class exported_component<T, true, sizeof(T), detail::sizeof_vptrs<T>::value> {
    char padding0[data_size/5];
    virtual void padding1() { throw std::invalid_argument("Not implemented"); }
    virtual void padding2() { throw std::invalid_argument("Not implemented"); }
    virtual void padding3() { throw std::invalid_argument("Not implemented"); }
    virtual void padding4() { throw std::invalid_argument("Not implemented"); }
    virtual void padding5() { throw std::invalid_argument("Not implemented"); }
  };
  template<typename T, bool resilent, size_t data_size> class exported_component<T, resilient, sizeof(T), 0> {
  };
  template<typename T, bool resilent, size_t data_size> class exported_component<T, true, sizeof(T), 0> {
    char padding0[data_size/5];
  };
}
\end{lstlisting}
\label{componentexport}
\end{program}

To export, we simply encode via template parameters into the tag type \texttt{exported\_component} the size of the data occupied by the exported type, plus how many vptrs it has (if it has virtual functions, you'd get one vptr, if you have virtual inheritance you'd get two, and so on). Note that \texttt{\_\_component\_export\_\_::<type>} gets the same symbol visibility as what it exports, so it appears in any DLL or shared object list of exported symbols. Note also if we're using the non-legacy method, we pad both the member variable and vtable with extra space.

The second part is a component linker program which you use to link your DLL/shared object instead of the standard linker: the component linker simply iterates the exported symbols list looking for any in the \texttt{\_\_component\_export\_\_} namespace. For each of those, it examines the return type of the \texttt{operator=()}, or if none, any \texttt{\_\_signature()} declared by that type by parsing the symbol mangling into a partial AST, thus extracting the original type being componentised, its size, whether it has virtual functions and inheritance and its declaration source file and line number for useful error printing. It then proceeds to generate a \emph{bindings declaration and implementation} which is simply a chunk of automatically generated C (not C++) code which lists out each of the member functions available from the componentised type with suitable C-compatible call stubs named after the MD5 of the signature of the symbol and type being exported, plus the order of declaration of the virtual member functions (which can be deduced by comparing the vtable contents to exported symbols, knowing that each component boundary's vtable will end with either a \texttt{\_\_end\_of\_vtable} or padding entry), plus the signatures of the exported functions (which again can be parsed out from their mangled symbols), with the stubs file compiled and linked into the shared library binary. Interestingly, the parameters of all exported functions are actually composed into input and output structures, this lets you easily work around vendor-specific weirdnesses in inverting calling conventions (I'm looking at you MSVC on x86!), plus allowing GCC with its Itanium ABI to call into the MSVC ABI and vice versa on Windows, plus it lets C++ metaprogramming entirely skip the C thunking layer and \emph{to inline direct calls into linked components} where safe e.g. on ARM, where all compilers always use the same ABI. Upon generation of the shared library which is now also a component, it can be imported into the graph database where the graph of API relationships can be reconstituted through examination of the component ABI metadata earlier bound into the shared library\footnote{Basically, a set of hidden functions are exposed which will return whatever metadata you want from the component. This allows easy versioning, plus lets you override those hidden functions with something customised for more specialised uses.}.

You now have a set of C-compatible bindings which let any C program instantiate any C++ object and use it -- why C-compatible? Because that is what almost every interpreted language uses e.g. Python, PHP, Lua etc. You'll note that the system employed is basically the same as that of COM, albeit somewhat modernised and more C++-specific. Unlike COM, we don't place any arbitrary restrictions on what kinds of C++ you can and can't let near the component ABI boundary -- rather, we let the link stage refuse to link if the underlying graph database says that some code modification will break the component's interface contract. This is, in my opinion, a better way of avoiding problems caused by implementation inheritance which was banned by Microsoft COM as being dangerous.

The third part of the example design deviates strongly from Microsoft COM, and this I call \emph{resilience wrapping} which very significantly mitigates the fragile base class and fragile binary interface problems mentioned earlier -- this works via the padding added by the metaprogramming to the member variable and vtable spaces. You can see how they work better through code than through English, so have a look at Program \ref{componentimport} which shows how we go about importing a component's interface for use.

\begin{program}[!t]
\centering
\caption{Component import}
\begin{lstlisting}[language=c++, morekeywords={}]
namespace __component_import_machinery__ {
  template<typename T, size_t data_size=sizeof(T), size_t vptrs_size=detail::sizeof_vptrs<T>::value> class imported_component : public T {
    char padding0[data_size/5];
    virtual void padding1() { throw std::invalid_argument("Not implemented"); }
    virtual void padding2() { throw std::invalid_argument("Not implemented"); }
    virtual void padding3() { throw std::invalid_argument("Not implemented"); }
    virtual void padding4() { throw std::invalid_argument("Not implemented"); }
    virtual void padding5() { throw std::invalid_argument("Not implemented"); }
  public:
    using T::T; 
  };
  template<typename T, size_t data_size> class imported_component<T, sizeof(T), 0> : public T {
    char padding0[data_size/5];
  public:
    using T::T; 
  };
}

namespace internal {
  namespace Boo {
    class Foo {
      int a;
      Foo(int _a) : a(_a) { }
    };
  }
}
namespace Boo {
  typedef __component_import_machinery__::imported_component<internal::Boo::Foo> Foo;
}
\end{lstlisting}
\label{componentimport}
\end{program}

This looks awfully similar to the exported component case, and that is because it is -- the reason why is because it is intended and expected that the same header file used to mark up a component for export is also used for import, so the idea is that very little changes just because you are importing -- in particular, everything looks the same in terms of namespace layouts, just that \texttt{exported\_component} is now an \texttt{imported\_component} such that the component linker can tell the difference between them.

This `resilience slosh factor' added to the member variable and vtable spaces is to enable the component linker to `cheat' when linking up slightly mismatching implementations with client code expecting a slightly different implementation. As much as you might think it would be fraught with dragons, that is exactly why the graph database is so essential: you can bundle into the graph (which can be updated from some central repository) which slight variations C++ component implementations are safe with which patterns of use code. The default rules are exceptionally conservative, but in fact the rules generation is very easy to automate with a bit of clang AST parsing (explained a little bit more next paragraph), so generally it all `just works'.

Unfortunately due to the need for brevity, the two remaining parts of the example design are those I'll describe least; the first, being how the link stage works, is very simple: the DLL/shared object is output just as before, and it still provides all the same shared library symbols it would as normal -- in other words, \emph{you can optionally link to a component as if it were a normal shared library} using the normal linker. This is done for backwards compatibility. If, however, you link using the component linker, those symbols which would link into implementation in an external shared library are instead linked to a delay resolving stub. This stub is able to ask the graph database during process bootstrap for the appropriate external component matching what the local component was compiled against, and if the hashes of their APIs match (i.e. it's a one to one match) a very fastpath link can be performed. If however they do not exactly match, one can parse through the exported component metadata, making use of the fact that member variable and vtables have been padded to allow a certain amount of slosh factor when linking up slightly incompatible components. If the slosh factor gets exceeded, or a breaking ABI change which cannot be resolved occurs, there are a number of interesting things which could be done: (i) the graph database could query an online master repo looking for a \emph{component adapter} which adapts one component's ABI into an older ABI (ii) is it possible that some ABI mismatches be safely worked around by \emph{auto-generating} a component adapter which simply exposes the available component's API to the old component's use patterns and if the standard C++ rules allow the compile (e.g. a programmer added a defaulted argument to the newer component, something which normally causes an ABI break but which is very easy to automatically resolve), simply insert the auto-generated component adapter between old and new versions (iii) for some really funky fun, one could even shim a thunk component built against some ancient implementation into a complete reimplementation of that component by doing on-the-spot type conversion as calls traverse the ABI layer\footnote{I tested one of these translating some toy classes like \texttt{std::string} between two totally incompatible STL implementations, and it worked better than expected.}. John Bandela's CppComponents implements exactly these incommensurate type implementation thunks for \texttt{std::string}, \texttt{std::vector}, \texttt{std::pair} and \texttt{std::tuple} such that one can mix libstdc++ from MinGW into a process using the Dinkumware STL from MSVC.

The last remaining part of the example design is how the event model is implemented, and I will say even less than about the link stage: the event model is fundamentally Boost.ASIO's asynchronous event dispatch and handling framework, with all other event models (e.g. Qt's event system) adapted into ASIO's using some unfortunately hacky code (I did always mean to get the component linker to patch out the use of the X11 or Win32 event model in component code with emulations written against ASIO, but I never got round to it as hacking Qt's source code was easier). I will say this though: given the popularity, flexibility and efficiency of ASIO's event handling model, it almost certainly would be the event model of choice of any future C++ object components implementation, and indeed the graph database design is entirely ASIO based.

Astute readers will have noticed that the link stage of this example can detect most kinds of interface break: reordering or modifying virtual functions, modifying non-virtual functions or any other form of signature breakage, adding or removing member variables. Note the one glaring exception: we cannot currently detect, neither via metaprogramming nor via examination of compiled objects, \emph{reordering}\footnote{By reordering I also include any changes which do not affect footprint i.e. \texttt{sizeof()}.} of member variables nor reordering base classes for classes without vtables. For those fragility and nasty surprises remain, until we can replace metaprogramming and mangled symbol parsing with a purely AST based solution (a hint as to a better, non-IP-tainted design for C++ object components!).

The key point being made in this example is that the \textbf{graph database} makes full fat C++ object components tractable -- while you could build a C++ component implementation without a graph database (and I have built one more than once now), you'll find that the thing getting in the way of scaling out such a solution to a standards-level global C++ industry-wide scale is the lack of a reliable, ACID-compliant, graph database which is low level enough to work at the process bootstrap level, because without a graph database the logic you must encode into the component loader to resolve all possible mismatches in component being loaded versus component compiled against is simply intractable, or at least it is without making the component objects implementation very brittle which defeats its purpose.

If you need more convincing and/or my word isn't good enough for you, try as a thought exercise building out an object components implementation using only SQLite3, remembering to include versioning, compatibility and shim adapters, avoiding any DLL hells, keeping performance good on copy-on-write filing systems, and coping sanely with ODR violation handling in a C++ modularised reflection-capable world. Oh, and you are doing this with just the C library and the STL before any shared libraries have been loaded, and it has to work right during early system bootstrap! Don't get me wrong: it can be done fairly straightforwardly with an ORM layer atop the SQL, but it's a lot harder, less flexible, and more brittle than with a freeform graph database, and I have first-hand knowledge that performance is not good. Besides, an embedded low level graph database is useful for storage and access of \emph{any} kind of structure of many bits of data, not just component objects, as we will see in the next section on Filesystem.

The dependency of graph databases on a component object implementation goes partially the other way, too: the potential of a graph database as proposed earlier can only be maximised if componentisation is available. This is because the proposed graph database is \emph{self bootstrapping} i.e. the code implementing the graph database is stored, versioned and loaded by the same graph database. When a process which uses the graph database first boots up, the component(s) implementing the graph database engine can be found by looking for the object with hash id 0x00000000000000000000000000000000 00000000000000000000000000000001 (hereafter \texttt{::1}) which will index a sequence of shared library implementations, architectures and versions which implement the graph database. Each one can be asked to examine the graph database to discover the correct engine version to use, thus bootstrapping the correct implementation for some given database, toolset, and architecture into memory. Upgrading the database is therefore a matter of successfully transacting a write of a new version of the engine binaries, kicking out all other readers and writers, unloading the old engine, loading in the new engine and asking it to perform the upgrade.

Similarly, the per-graph indexers such as Boost.Graph would also be best implemented exactly the same way: the component indexer which has hash id \texttt{::1} might have the Boost.Graph indexer at hash id \texttt{::2}, the remote fetcher of indexers from the internet indexer at hash id \texttt{::3}, and so on. Something I haven't detailed here at all is the graph database security model such that graphs can have arbitrary per-graph security models, well for that you would very definitely need a components implementation. My ideal outcome is that the proposed graph database is merely the first step in eventually actualising that shard of a distributed global active object database from my aforementioned next-generation platform, but that is several decades away yet at the current pace of work.

And C++ object components are but one of the many things a standardised core C++ graph database makes possible -- let us turn quickly to transcending Filesystem.

\subsection{Notes on implementing Object Components under an `as if everything header only' model}

Clearly object components in C++ have three main parts: (i) a rigid but loosely coupled ABI (ii) a memory management/lifetime management model and (iii) an (asynchronous) event model. As mentioned in the previous discussion, the form of the latter two are fairly obvious for C++ object components, but a GLSL shader-esque per-type source compilation model which potentially outputs lots of little C++ Modules is quite different.

I think that such a vision actually makes an even stronger case for object components in C++, because you're going to need some method of corralling so many C++ Modules into change resilient groups, and I don't think Module partitions go far enough, especially as for very large code bases you need to draw hard boundaries to contain the change ripple detection mechanism going nuts as you try to compile code. Basically, you need a mechanism for saying that `this bunch of code is to be treated for as-if header only purposes as a single opaque block providing the following rigid unchanging ABI. Any change ripples landing here are to be reflected back at the sender'.

One thing is for sure though: with a change ripple management mechanism in place, object components look \emph{utterly} different than to traditional software components. In particular, I think that the ABI of such software components must be active i.e. they can be programmed to actively respond to incoming change ripples in some bespoke fashion rather than necessarily imposing a fixed ABI. The idea here would be to still further loosen the coupling between components in a way not possible before.

As I hinted at earlier, the `Concept checks for the ABI' are ideal for programmatically implementing a software component's rigid but loosely coupled ABI, and I could see it being possible that `components' are merely a database-enforced abstraction that has no basis outside the database making it so.

As for the remaining parts of lifetime management and event model, well the proposed graph database will need to implement those anyway as part of being implementation self-bootstrapping as earlier described. My current expectation is to use John Bandela's CppComponents for that part, but with additional requirements.

\section{Second potential killer application for a C++ Graph database: Filesystem}

I did mention at the very outset in the abstract that I would be proposing to tie together SG2 Modules and SG7 Reflection with SG3 Filesystem, so what did I mean by tying in Filesystem?

I left this to last because I find it an ongoing amazement how hard it is in 2014 to write to more than one file from more than one process in a reliable way (i.e. power loss and race condition safe) -- in fact, we have actually gone \emph{backwards} compared to the systems of the 1970s in writing code which reliably shepherds program state over long periods of time, and I think a lot of why this has happened was due to the highly visible failures of next generation filing systems which were supposed to have given us a database filing system on every computing system some years ago. With such visible cancellations\footnote{Microsoft's third major attempt at a database filing system after a decade of failures was called WinFS and it succeeded the prior attempts Object Filing System and Relational Filing System. WinFS is probably the most famous example of a `failed' database filing system because it actually worked, but it imposed performance consequences on certain work loads which were at the time considered a showstopper. Much of WinFS ended up being folded into SQL Server, which shows that people want database features in databases, not in filing systems.}, I think database filing systems were marked down as being `too hard', and R\&D budgets were allocated elsewhere.

We now know that filing systems aren't going to significantly exceed the copy-on-write B-tree designs used by ZFS, Btrfs and Microsoft's ReFS -- and don't get me wrong, they are an enormous improvement over the present universally available design which are journaling filesystems, but they are still no database.

And nor, do I think, should they attempt to be one. A \emph{lot} of users of filing systems have zero interest in even persisting the data written to files -- the filing system is basically being used as a global namespace through which processes coordinate. Somewhere inbetween are files which you'd like to keep around for a while, but aren't terribly important -- scratch files, anything which can be regenerated easily, that sort of thing. After that comes files which are extremely important and must not be lost, but can tolerate an occasional bit flip (video, music etc). And finally there are files where a single bit flip is serious: any executable file, or perhaps your company accounts or the sole copy of your SSH master private key for example.

The point here is that \emph{we use one filing system for many things} and using the wrong filing system for the task at hand is suboptimal, which is probably why database filing systems withered on the vine. What is proposed by this position paper is therefore rather exciting: \emph{the ability to apply as much or as little `database' to a filing system as needed depending on the use case}.

Because the proposed design is capable of operating at process bootstrap level, you can just go ahead and build it straight into a post-C++ 17 \texttt{std::filesystem} API as a pseudo-filing system namespace. Some months ago an interesting proposal by Alexander Lamaison came onto boost-devs about making Filesystem generic: you can see the proposal at \url{http://alamaison.github.io/2014/01/09/proposal-generic-filesystem/} and the boost-devs discussion at \url{http://goo.gl/Qzry4c}. Generic Filesystem was intended mainly for allowing things like ZIP archives to be treatable as just another filesystem space just as you would take for granted in Python \emph{et al}, but the system proposed could be very useful for overlaying different amounts of `database-ness' onto the filing system using the proposed graph database.

For example, let us say that a software project keeps three types of data: (i) runtime logs (ii) personally identifying and private information of a kind you must steward carefully due to legislation e.g. personal health information and (iii) security audit logs for whom accesses what and when -- you could imagine a health insurance application might do this for example. A contemporary design might store the runtime logs in some logs directory in \texttt{/tmp} and the personal health information and security audit logs in a SQLite3 database, and such a design would probably be considered minimally sufficient. But consider a design based on the proposed graph database: firstly, all three types of information now live in the same graph store, but the logs graph is marked as not particularly important, the personal information graph is stored as tier 2 data encrypted with each person's records being only accessible through an active database object component which tags and iterates the version of the personal data with every access, thus making every access to personal data permanently recorded both as part of the personal data graph and the global access graph. Because the graph store uses content hashes to encode history just as the git algorithm does, now the only way of deleting the fact someone accessed some person's records is to delete all history back to just before the access, and with that the design is now at the state of the art in security and resilience with very little work on the part of the C++ developer.

Best of all, the database of personal information can appear to C++ code using Filesystem as just another set of files living in directories, or even as a set of emulated SQL tables such that the previous SQLite3 code can be mostly reused without modification -- you could update the data via either route and whichever view is always up to date. I think this vision is very compelling myself.

\section{Conclusions}

I have hopefully shown fairly uncontroversially how some of C++'s greatest strengths impose unique consequences on C++ programmers. We have seen how I believe that the era of free lunching on throwing ever more C++ code into each compilation unit as a way of working around many of those unique-to-C++ issues will shortly taper out on present toolset architectures, and we will in my opinion once again face managing large numbers of separately compiled bits of C++ in a way more reminiscent of the 1990s than the past two decades. Because we have evolved so few new methods and processes of managing C++ ABI since the late 1990s, we are facing a possibility that most of the advances in C++ of the past twenty years will not be available to us when approaching ABI boundaries, something already taken for granted as being true by almost any very large scale C++ user.

In reviewing these scalability and complexity management problems, I have tried to show how C++ compiler vendors will almost certainly have to implement their own graph databases and/or extensions to the graph databases they are already using. I proposed that instead of each compiler vendor reinventing the wheel, the Boost C++ libraries could gain an implementation of a very generic, very extensible and very low level lightweight graph database with per-graph indexing implementations (e.g. Boost.Graph or SQLite3 could be specified as the index implementation for some graphs, but not others). I then went on to demonstrate some of the interesting things you could do with such a graph store at the core of the C++ runtime, including C++ as if everything were being compiled header only, active database objects/C++ components, multiple Filesystem spaces of differing qualities, and a central, portable and common process for coordinating the outputs from any C++ Modules and Reflection implementations.

For most readers, the final question will surely be `how feasible is all of this?' To implement the generic ACID-capable transactional content-addressable persistence layer alone (and on top of which you could fairly easily add arbitrary per-graph indexers), you need the following prerequisites:

\begin{enumerate}
\item A method of parallel hashing many pieces of content simultaneously. If you examine the simplified git algorithm described earlier, you will notice that even a single byte modification of a data item requires hashing at least two files and writing at least three files, with transactions potentially increasing the number of hashings by between O(N) and O(N log N) where N is the number of files modified in a transaction.
\item A platform abstraction layer to enable portable asynchronous file i/o. As everything happens in files, we will read and write an awful lot of separate files at once during our normal course of operation. Therefore, the ability to execute batches of asynchronous file i/o would be a huge help.
\item A platform abstraction layer to enable portable asynchronous file locking\footnote{Getting this right is unfortunately very hard on POSIX -- see \url{http://0pointer.de/blog/projects/locking.html} for how badly broken file locks are.} and copy-on-write delta file copies\footnote{The copy-on-write filing systems usually allow you to make cheap copies of files, storing only the extents which you modify.}.
\end{enumerate}

That is, of course, a minimal set of prerequisites -- one could improve performance hugely if you could use a shared memory region instead of lock files for example, but all that can be added later and besides, as I mentioned earlier, I don't think that write transaction performance is actually all that important compared to the other major benefits.

Well, I can tell you that progress on the prerequisites is going well: for Google Summer of Code 2013 myself and Paul Kirth (who is also presenting at this conference) successfully submitted Boost.AFIO, a portability layer for asynchronous file i/o based on Boost.ASIO, for Boost peer review. I am about half way through a SIMD threaded content hasher which can execute sixteen SHA-256's in parallel on a quad core Intel or ARMv7 CPU, yielding already a 1.2 cycle per byte average on Intel which is phenomenal. I am not happy with the design of the present implementation however, I believe much more of it could be metaprogrammed with optimisation hints as the current design defeats all C++ compiler optimisers. Once this conference is done, I am entirely intending to dedicate what spare time I get to finishing the content hasher, and once done to move onto the portability layer for asynchronous file locking.

Of course, one can only do so much in one's free time, so if any of the above excites you, here are some options available to help out the effort:
\begin{itemize}
\item I have a sore need for comprehensive functional testing of exception safety in proposed Boost.AFIO -- it is the only part not properly unit tested to destruction, yet handling errors perfectly is paramount for any code handling people's data. I am fairly confident it \textbf{is} exception safe, but I know it can leak resources sometimes in some exception throw routes, and all those need to be found and fixed with an exception safety test framework needed for the CI server to run per-commit to ensure exception safety remains perfect.

A decent exception safety unit test framework for Boost would actually be a great project in and of itself -- Boost.Test doesn't do much for exception safety testing (or at least to my knowledge), especially if it could hook in with \texttt{valgrind} (or whatever) to also track resource leaks.

\item The amount of testing involved in writing a cross-platform, network filing system aware, asynchronous file locking library for Boost is immense, especially getting a Jenkins CI configured with all the possible use case scenarios e.g. BSD NFS talking to Linux NFS which exports a Windows Samba share. I think that the writing of such a library is probably not that hard, but the testing will be the main time consumer. While the code is not written at the time of writing this paper, maybe by the time you are reading this now it is, if so I could do with help configuring test scenarios and getting all those hooked into a CI for per-commit testing as well.

\item Paul Kirth is still working on a fast, scalable portable asynchronous directory change monitor which we'll need for implementing concurrent database transactions and garbage collection, and I am sure he would appreciate some help. It actually is far harder than you might think -- he'll talk during his presentation at this conference on just how hard it is.

\item Proposed Boost.AFIO remains in the community review queue, and I would expect it will need several rounds of peer review before it could be recommended for acceptance into Boost. A review manager willing to invest the work involved would be hugely appreciated.

\item Financial sponsorship would make a huge difference to rate of progress as it would enable me (I have my own consulting firm) and other members of the Boost community to spend more hours on writing, testing and documenting the prerequisites for an embedded C++ graph database instead of being consumed by other revenue generating work. If you think any of the proposed features described above would be really useful to your employer, please do hook me up with them!
\end{itemize}

\section{Acknowledgements}
I am particularly indebted to Bjorn Reese who has given so freely of his time in reviewing proposed Boost.AFIO for various warts and omitted pieces. AFIO is considerably a better library for it. He also pointed out an entire missing section in a draft of this paper which I also somehow had mentally omitted, and which required another thousand words to fix -- I am terribly grateful.

Particular thanks go to reviewers Andrew Sutton and Stephan T. Lavavej for their insights and corrections, and without whose dialogue I would not have realised that I was actually asking for a C++ toolset where all code is compiled as if it were header only. Thanks are also due to Antony Polukhin who wrote no less than three sets of notes on a draft of this paper, as well as to Fraser Hutchison whose eagle eyes spotted a remarkable number of typos considering how many revisions this paper had already undergone. My thanks to Paul Kirth for working so hard during GSoC 2013 to bring AFIO into a state fit to enter the Boost peer review queue, I couldn't have done it without you.

Finally, I am especially grateful to my significant other, Megan, who has tolerated with surprising good humour the non-stop sixty plus hour weeks caused by these outside of work side projects. Such a life is particularly hard on family and loved ones.

\section{About the author}
Niall Douglas is one of the authors of proposed Boost.AFIO and is currently the primary Google Summer of Code administrator for Boost. He is an Affiliate Researcher with the Waterloo Research Institute for Complexity and Innovation at the University of Waterloo, Canada, and holds postgraduate qualifications in Business Information Systems and Educational and Social Research as well as a second undergraduate degree double majoring in Economics and Management. He has been using Boost since 2002 and was the ISO SC22 (Programming Languages) mirror convenor for the Republic of Ireland 2011-2012. He formerly worked for BlackBerry 2012-2013 in their Platform Development group, and was formerly the Chief Software Architect of the Fuel and Hydraulic Test Benches of the EuroFighter defence aircraft. He is a published author in the field of Economics and Power Relations, is the Social Media Coordinator for the World Economics Association and his particular interest lies in productivity, the causes of productivity and the organisational scaling constraints which inhibit productivity.

\bibliographystyle{abbrv}
\bibliography{references}

\begin{thebibliography}{10}

\bibitem{cernreflex}
Cern reflex.
\newblock \url{http://root.cern.ch/drupal/content/reflex}.

\bibitem{modulesinclang}
clang modules documentation.
\newblock \url{http://clang.llvm.org/docs/Modules.html}.

\bibitem{boostmirror}
Proposed boost.mirror.
\newblock \url{http://kifri.fri.uniza.sk/~chochlik/mirror-lib/html/}.

\bibitem{SSDsVsHardDrives}
Ssds vs hard drives 1980-2014 raw data.
\newblock \url{http://nedprod.com/studystuff/SSDsVsHardDrives.xlsx}.

\bibitem{n3340}
D.~M. Berris, M.~Austern, and L.~Crowl.
\newblock N3340: Rich pointers.
\newblock {\em Proceedings of the ISO JTC1/SC22/WG21 The C++ Programming
  Language committee}, 2012.
\newblock
  \url{http://www.open-std.org/jtc1/sc22/wg21/docs/papers/2012/n3340.pdf}.

\bibitem{box1998essential}
D.~Box.
\newblock {\em Essential Com}.
\newblock Addison-Wesley Professional, 1998.

\bibitem{spectrum2013}
R.~Courtland.
\newblock The status of moore's law: It's complicated.
\newblock {\em IEEE Spectrum}, 2013.
\newblock
  \url{http://spectrum.ieee.org/semiconductors/devices/the-status-of-moores-law-its-complicated}.

\bibitem{ganapathi2004pcs}
A.~Ganapathi, Y.-M. Wang, N.~Lao, and J.-R. Wen.
\newblock Why pcs are fragile and what we can do about it: A study of windows
  registry problems.
\newblock In {\em Dependable Systems and Networks, 2004 International
  Conference on}, pages 561--566. IEEE, 2004.
\newblock
  \url{https://www.cs.cmu.edu/~nlao/publication/older/ganapathia_DSN_registry.pdf}.

\bibitem{kuhn1962structure}
T.~S. Kuhn.
\newblock {\em The structure of scientific revolutions}.
\newblock University of Chicago press, 1962.

\bibitem{sutter2005free}
H.~Sutter.
\newblock The free lunch is over: A fundamental turn toward concurrency in
  software.
\newblock {\em Dr. Dobb's Journal}, 30(3):202--210, 2005.
\newblock
  \url{http://www.mscs.mu.edu/~rge/cosc2200/homework-fall2013/Readings/FreeLunchIsOver.pdf};
  2009 updated edition
  \url{http://www.gotw.ca/publications/concurrency-ddj.htm}.

\bibitem{n1426}
H.~Sutter and T.~Plum.
\newblock N1426: Why we can't afford export.
\newblock {\em Proceedings of the ISO JTC1/SC22/WG21 The C++ Programming
  Language committee}, 2003.
\newblock
  \url{http://www.open-std.org/jtc1/sc22/wg21/docs/papers/2003/n1426.pdf}.

\bibitem{n3889}
A.~Sutton.
\newblock N3889: Concepts lite technical specification (early draft).
\newblock {\em Proceedings of the ISO JTC1/SC22/WG21 The C++ Programming
  Language committee}, 2014.
\newblock
  \url{http://www.open-std.org/jtc1/sc22/wg21/docs/papers/2014/n3889.pdf}.

\bibitem{n3347}
D.~Vandevoorde.
\newblock N3347: Modules in c++ (revision 6).
\newblock {\em Proceedings of the ISO JTC1/SC22/WG21 The C++ Programming
  Language committee}, 2012.
\newblock
  \url{http://www.open-std.org/jtc1/sc22/wg21/docs/papers/2012/n3347.pdf}.

\bibitem{williams1990inheritance}
T.~Williams.
\newblock On inheritance, what it means and how to use it.
\newblock {\em Draft, Applications Architecture Group, Microsoft Research},
  1990.

\end{thebibliography}

\end{document}